\documentclass[reprint,twocolumn,prb,amsmath,amssymb,superscriptaddress,notitlepage,nofootinbib,longbibliography]{revtex4-1}

\usepackage{xcolor}
\usepackage[export]{adjustbox}
\usepackage{lipsum}
\usepackage{hyperref}
\usepackage{graphicx}  
\usepackage{amsmath, amsfonts, amssymb}
\usepackage{braket}
\usepackage{dsfont}
\usepackage{upgreek}
\usepackage{qcircuit}
\usepackage{tikz}
\usetikzlibrary{matrix}

\usepackage[utf8]{inputenc}
\usepackage[english]{babel}
\usepackage{hyperref}
\usepackage{array}
\usepackage{mathtools}
\usepackage{booktabs}
\usepackage{siunitx}

\newcommand{\CW}{\circlearrowright}
\newcommand{\CCW}{\circlearrowleft}

\newcommand{\id}{\mathds{1}}

\newcommand{\ketbra}[2]{
    \lvert #1 \rangle \! \langle #2 \rvert
}

\newcommand{\cz}{\text{c}\sigma_{z}}

\newcommand{\smat}[4]{
    \scriptsize
    \begin{pmatrix}
    #1 & #2 \\
    #3 & #4
    \end{pmatrix}
}

\newcolumntype{C}{>{$}c<{$}}
\AtBeginDocument{
    \heavyrulewidth=.08em
    \lightrulewidth=.05em
    \cmidrulewidth=.03em
    \belowrulesep=.65ex
    \belowbottomsep=0pt
    \aboverulesep=.4ex
    \abovetopsep=0pt
    \cmidrulesep=\doublerulesep
    \cmidrulekern=.5em
    \defaultaddspace=.5em
}

\begin{document}

\title{Deterministic photonic quantum computation in a synthetic time dimension}

\author{Ben Bartlett}
\email{benbartlett@stanford.edu}
\affiliation{Department of Applied Physics, Stanford University, Stanford, California 94305, USA}
\affiliation{Department of Electrical Engineering, Stanford University, Stanford, California 94305, USA}

\author{Avik Dutt}
\affiliation{Department of Electrical Engineering, Stanford University, Stanford, California 94305, USA}

\author{Shanhui Fan}
\email{shanhui@stanford.edu}
\affiliation{Department of Electrical Engineering, Stanford University, Stanford, California 94305, USA}

\begin{abstract}
Photonics offers unique advantages as a substrate for quantum information processing, but imposes fundamental scalability challenges. Nondeterministic schemes impose massive resource overheads, while deterministic schemes require prohibitively many identical quantum emitters to realize sizeable quantum circuits. Here we propose a scalable architecture for a photonic quantum computer which needs minimal quantum resources to implement any quantum circuit: a single coherently controlled atom. Optical switches endow a photonic quantum state with a synthetic time dimension by modulating photon-atom couplings. Quantum operations applied to the atomic qubit can be teleported onto the photonic qubits via projective measurement, and arbitrary quantum circuits can be compiled into a sequence of these teleported operators. This design negates the need for many identical quantum emitters to be integrated into a photonic circuit and allows effective all-to-all connectivity between photonic qubits. The proposed device has a machine size which is independent of quantum circuit depth, does not require single-photon detectors, operates deterministically, and is robust to experimental imperfections.
\end{abstract}

\maketitle

\section{Introduction}
Photonics offers many advantages for quantum information processing~\cite{Wang2020IntegratedTechnologies, obrien_optical_2007, Zhong2020QuantumPhotons}: optical qubits have very long coherence times, are maintainable at room temperature, and are optimal for quantum communication. The main difficulty faced by all quantum computing (QC) architectures is scalability, but this is especially true for photonic systems. Optical qubits must propagate, so processing must be done mid-flight by passing the photons through sequential optical components. Since photonic quantum gates are physical objects (as opposed to, e.g. sequential laser pulses for atomic qubits), machine size scales with circuit depth, making complex quantum circuits prohibitively large to implement even using compact integrated photonics.

Further limiting the scalability of photonic quantum computers is the difficulty of integrating many high-fidelity multi-photon gates into an optical circuit. This is an issue both for nondeterministic gate schemes~\cite{Knill2001AOptics, Kok2007LinearQubits}, which impose massive resource overheads for fault tolerant operation due to low gate success probabilities~\cite{Li2015ResourceComputing}, and for deterministic scattering-based approaches~\cite{tiecke_nanophotonic_2014, turchette_measurement_1995, Duan2004ScalableInteractions, Reiserer2015Cavity-basedPhotons, Zheng2013Waveguide-QED-basedComputation}. Although scattering-based two-photon gates can be individually implemented with high fidelity~\cite{hacker_photonphoton_2016, tiarks_photonphoton_2019, volz_nonlinear_2014, beck_large_2016, fushman_controlled_2008}, unrealistically large numbers of identical quantum emitters are needed to realize sizeable quantum circuits~\cite{Bartlett2020UniversalProcessing}, a problem which is exacerbated in solid-state quantum emitters by poor indistinguishability due to homogeneous and inhomogeneous broadening~\cite{machielse_quantum_2019, aharonovich_solid-state_2016}. An architecture for a quantum computer which uses only a single quantum emitter to implement all gates in a quantum circuit would thus substantially improve the scalability and experimental feasibility of scattering-based photonic quantum computation.

\begin{figure*}[t]
\centering 
\includegraphics[width=\textwidth]{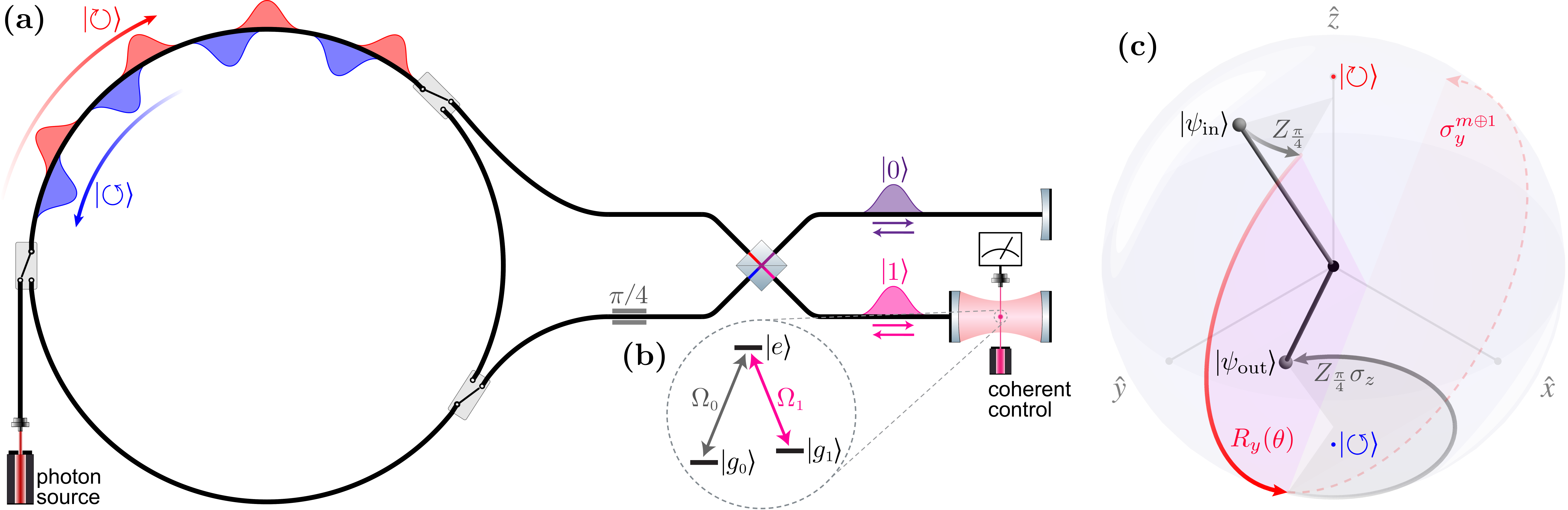}
\caption{
The photonic quantum computer architecture described in this work.
\textbf{(a)} 
The physical design of the device. Photonic qubits counter-propagate through a fiber storage ring and optical switches can selectively direct photons through a scattering unit to interact with an atom in a cavity which is coherently controlled by a laser. 
\textbf{(b)} 
The energy structure of the atom: $\Omega_1$ is resonant with the cavity mode and photon carrier frequency, while $\Omega_0$ is far-detuned.
\textbf{(c)} 
Bloch sphere depiction of the state of a photonic qubit in the $\left\{ \ket{\CW}, \ket{\CCW} \right\}$ basis and an operation applied by one pass through the scattering unit. The rotations about $\hat z$ by fixed angles (grey) are applied by the phase shifter and beamsplitter, while the rotation about $\hat y$ by a controllable angle $\theta$ (solid red) is applied to the atom using the cavity laser. Projectively measuring the atom teleports this rotation onto the photon, but may overshoot the target angle $\theta$ by $\pi$ (dotted red) depending on the measurement outcome $m$. This operation is a universal single-qubit primitive: by composing several of these operations and adapting subsequent rotation angles based on measurement outcomes, arbitrary single-qubit gates can be deterministically constructed.
}
\label{fig:ring_computer}
\end{figure*}

Here we show that the emerging concept of synthetic dimensions~\cite{yuan_synthetic_2018, boada_quantum_2012} naturally lends itself to such an architecture. Synthetic dimensions have recently generated great interest for exploring topological physics in photonics~\cite{ozawa_topological_2019}, but have not been extensively applied to quantum photonic systems. To form a synthetic dimension, one designs the couplings between states of a system, either by repurposing the usual geometric dimensions, such as space\cite{lustig_photonic_2019} or time~\cite{regensburger_paritytime_2012, wimmer_experimental_2017, marandi_network_2014, inagaki_coherent_2016, mcmahon_fully_2016}, or by augmenting these dimensions with internal degrees of freedom, such as frequency~\cite{yuan_photonic_2016, ozawa_synthetic_2016, bell_spectral_2017, reimer_high-dimensional_2019, yang_mode-locked_2020, wang_multidimensional_2020}, spin~\cite{celi_synthetic_2014, mancini_observation_2015, stuhl_visualizing_2015, dutt_single_2020}, orbital angular momentum~\cite{luo_quantum_2015, yuan_photonic_2019}, or Floquet-induced side bands~\cite{baum_setting_2018, martin_topological_2017}. Since couplings between states within the synthetic dimension can be dynamically reconfigured and are not fixed by physical structure, one can scalably implement lattices with intricate connectivity. This allows multiple photonic qubits to be manipulated in synthetic space by a single quantum emitter without requiring spatially separated structures.

Our proposed design consists of a fiber ring coupled to a cavity containing a single coherently controlled atomic qubit. Optical switches endow the counter-circulating photonic states with a synthetic temporal dimension by allowing coupling between these states. By scattering photons against the atom and subsequently rotating and projectively measuring the atomic state, operations can be teleported onto the photonic qubits; these operations can be composed to deterministically construct any quantum circuit. Readout of the photonic quantum state can be performed without the need for single-photon detectors by sequentially swapping the state of the atom with each photonic qubit.

Our scheme has several unique characteristics. Most notably, the only controllable quantum resource is the single atomic qubit. All quantum~\cite{Note1} 
manipulations and measurements of the photonic qubits are carried out indirectly by operations performed on this atom and teleported to the photons.
This reduces the primary implementation challenge to preparing a single strongly coupled atom-cavity system, which has been experimentally demonstrated many times~\cite{colombe_strong_2007, gehr_cavity-based_2010, samutpraphoot_strong_2020, mckeever_experimental_2003, covey_telecom-band_2019, fushman_controlled_2008, Reiserer2015Cavity-basedPhotons, chang_colloquium_2018}.
The synthetic time dimension allows the single atom to serve as the nonlinearity for all quantum gates and provides effective all-to-all connectivity between the photonic qubits. The programmable nature of the teleported gates allows the atom to sequentially apply each required single- and two-photon gate without complex photon routing. This negates the requirement of conventional photonic QC schemes for many identical quantum emitters to be integrated into a photonic circuit. Finally, this design does not require single-photon detectors, which are a significant limitation to photonic QC. Measurement of the atomic state can be performed with near-$100\%$ efficiency using the quantum jump technique, greatly improving the scalability of this design~\cite{Monroe2002QuantumCavities, Duan2004ScalableInteractions, gehr_cavity-based_2010}. 

Related to but distinct from this work are proposals for generating time- and frequency-multiplexed 2D cluster states using a single or pair of quantum emitters~\cite{pichler_universal_2017, economou_optically_2010} and experimental demonstrations using parametric nonlinearities~\cite{asavanant_generation_2019, larsen_deterministic_2019}. Although 2D cluster states are a universal resource for quantum computation \cite{Nielsen2006Cluster-stateComputation} when combined with measurement,
the schemes that prepare these states only apply a single type of quantum operation to the photonic qubits, and photon detectors with their associated limitations are required for further processing and state readout. In contrast, our scheme directly implements the quantum circuit model of QC, can deterministically construct any quantum gate, and can perform state readout without the need for photon detectors.


\section{Design}
The architecture for the scheme is shown in Figure \ref{fig:ring_computer}(a). Qubits are encoded as trains of single photon pulses counter-propagating through an optical storage ring, where the two propagation directions $\left\{ \ket{\CW}, \ket{\CCW} \right\}$ form the computational basis. A single-photon source injects photon pulses into the ring; each photon is spectrally narrow about a carrier frequency $\omega_c$, has a pulse width $\tau$, and occupies its own time bin with temporal spacing $\Delta t \gg \tau$. (The photon source need not be deterministic as long as the time bin of each photon can be resolved. Alternately, a dedicated single-photon source may not be needed, as the atom-cavity system discussed below could itself be used as the source by using the control laser to excite the atom~\cite{McKeever2004DeterministicCavity,Duan2003CavityAtoms}.)

The storage ring contains a pair of asymmetrically placed~\cite{Note2}
optical switches, which can selectively direct photons from the ring through a static 50:50 beamsplitter and $\pi/4$ phase shifter and into a pair of waveguides. One of these waveguides is coupled to a cavity containing a single atom with a $\Lambda$-shaped three-level energy structure, shown in Figure \ref{fig:ring_computer}(b). The atom has non-degenerate ground states $\ket{g_0}$ and $\ket{g_1}$ and an excited state $\ket{e}$, and the $\ket{g_1} \leftrightarrow \ket{e}$ transition at frequency $\Omega_1$ is resonant with cavity mode frequency and photon carrier frequency $\omega_c$. The atom is coherently controlled by a laser which applies rotations between $\ket{g_0}$ and $\ket{g_1}$, and its state can be measured in the $\left\{\ket{g_0}, \ket{g_1}\right\}$ basis. We refer to the subsystem consisting of everything except the storage ring and photon source (the right half of Figure \ref{fig:ring_computer}(a)) as the ``scattering unit''. The round-trip optical path length through the scattering unit is matched to the path length around the storage ring so that a photon returns to its original time bin after passing through the scattering unit.


After a photon scatters against the atom and is returned to the storage ring, a rotation is applied to the state of the atomic qubit and a projective measurement is performed, teleporting the rotation onto the photonic qubit, as shown in the next section. By composing three of these teleported rotations, arbitrary single-qubit gates can be deterministically constructed. A controlled phase-flip ($\cz$) gate between two photons can also be constructed with a similar process, enabling universal quantum computation. Readout of the final quantum state can be performed without the need for single-photon detectors by sequentially swapping the state of the atom with each photonic qubit.



\subsection{Rotation teleportation mechanism}
Here we outline the mechanism by which a rotation gate may be teleported onto a photonic qubit; we show in the next section that by composing these teleported rotations, arbitrary single-qubit gates may be constructed. A derivation of the mechanism described here is shown in greater detail in the Supplementary Information.~\cite{SM} Suppose we wish to apply a rotation to photon $j$, which occupies time bin $t_j$ and is circulating in the storage ring in state $\ket{\psi_\text{in}} = \alpha \ket{\CW} + \beta \ket{\CCW}$, where $\ket{\CW}$ and $\ket{\CCW}$ denote the two counter-circulating states. While the optical switches are in the ``closed'' state, photons remain inside the storage ring; to operate on photon $j$, we ``open'' the switches at time $t_j - \Delta t / 2$ and close them again at $t_j + \Delta t / 2$ to direct photon $j$ into the scattering unit. The photon passes through a $\pi/4$ phase shifter, which applies (up to a global phase) a $Z_\frac{\pi}{4} \equiv R_z (\pi/4) = \smat{e^{-i\pi/8}}{0}{0}{e^{i\pi/8}}$ rotation, and a 50:50 beamsplitter, which applies $B = \frac{1}{\sqrt{2}} \smat{1}{i}{i}{1}$. Before interacting with the atom, the photon is a superposition of modes in the top and bottom waveguides; we label these spatial modes as $\ket{0}$ and $\ket{1}$, respectively. We can thus relate the basis states of the ring and scattering unit via the unitary transformation $\left\{ \ket{0}, \ket{1} \right\} = B Z_\frac{\pi}{4} \left\{ \ket{\CW}, \ket{\CCW} \right\}$.

The $\ket{0}$ component of the photon state is reflected by a mirror in the top waveguide, imparting a $\pi$ phase shift, while the $\ket{1}$ component undergoes a cavity-assisted interaction with the atom in the bottom waveguide, which is initialized in the state $\ket{+} \equiv \frac{1}{\sqrt{2}} \left(\ket{g_0} + \ket{g_1}\right)$. The $\ket{g_1} \leftrightarrow \ket{e}$ transition frequency $\Omega_1$ is resonant with the cavity mode and photon frequency $\omega_c$, while the $\ket{g_0}\leftrightarrow\ket{e}$ frequency $\Omega_0$ is far-detuned. Thus, relative to the phase of the $\ket{0}$ mode, a $\pi$ phase shift is applied to the $\ket{1}\otimes\ket{g_1}$ component of the $\ket{\text{photon}} \otimes \ket{\text{atom}}$ quantum state, implementing the unitary transformation corresponding to a controlled-Z gate between the atom and the photon, $\cz = e^{i \pi \ketbra{1}{1} \otimes \ketbra{g_1}{g_1}}$. After scattering, the photon passes back through the beamsplitter and phase shifter and is returned to the storage ring. The joint state $\ket{\Phi}$ of the photon-atom system after a round trip through the scattering unit is:
\begin{equation}
\label{eq:phi_pre_measurement}
\ket{\Phi} = \left( Z_\frac{\pi}{4} B \otimes \id \right) \cz \left( B Z_\frac{\pi}{4} \otimes \id \right) \left( \ket{\psi_\text{in}} \otimes \ket{+} \right).
\end{equation}

\begin{figure}[t]
\centering 
\includegraphics[width=\columnwidth]{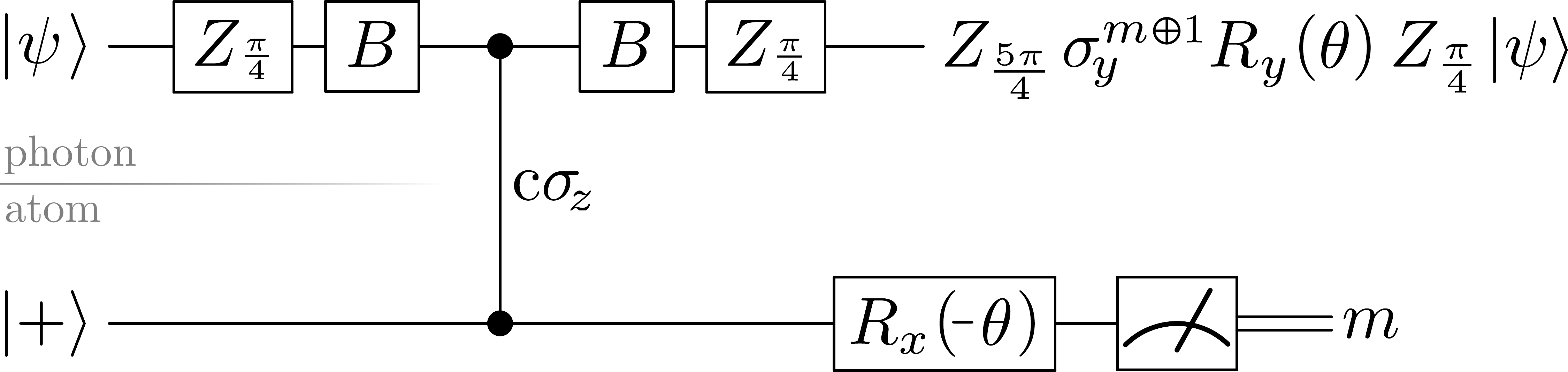}
\caption{Quantum gate sequence corresponding to one pass of a photon through the scattering unit. The projective measurement teleports the rotation applied to the atomic qubit onto the photonic qubit.}
\label{fig:gate_sequence}
\end{figure}

After the photon has returned to the storage ring, a rotation $R_x\left( -\theta \right) = \exp\left( i \sigma_x \theta/2 \right)$ is applied to the atomic qubit.
Finally, a projective measurement of the atomic state in the $\left\{\ket{g_0}, \ket{g_1}\right\}$ basis is performed, obtaining a bit $m\in\{0,1\}$. As shown in the Supplementary Information  \cite{SM}, this atomic measurement projects the state of the photonic qubit to:
\begin{align}
\begin{split}
\label{eq:psi_out}
\ket{\psi_\text{out}} &= Z_{\frac{\pi}{4}} \sigma_z \left(-\sigma_y \right)^{m\oplus 1} R_y\left(\theta\right) Z_{\frac{\pi}{4}} \ket{\psi_\text{in}} \\
&= i^m Z_{\frac{5 \pi}{4}} R_y\left( \theta + \pi (m \oplus 1) \right) Z_{\frac{\pi}{4}} \ket{\psi_\text{in}},
\end{split}
\end{align}
where $R_y \left( \theta \right) = \exp\left( - i \sigma_y \theta/2 \right)$ and $m \oplus 1$ denotes addition modulo 2. Thus, the measurement teleports the $R_x(\theta)$ rotation of the atom to the $R_y(\theta)$ or $R_y (\theta+\pi)$ rotation of the photon, depending on $m$. The full sequence of operations is shown in Figure \ref{fig:gate_sequence}. 

This teleportation scheme is an inversion of the paradigm of teleportation-based quantum computing \cite{Jozsa2005AnComputation, Gottesman1999QuantumPrimitive, Nielsen2006Cluster-stateComputation}: in both cases, the original data qubit is entangled with an ancilla using a $\cz$ operation, but instead of rotating and measuring the data qubit to teleport the modified state onto the ancilla, in our scheme we rotate and measure the ancilla (the atom) to teleport a rotation onto the data qubit (the photon).

\subsection{Constructing arbitrary single-qubit gates}
We now show that the teleported gate operation of Eq. \ref{eq:psi_out} is sufficient to construct arbitrary single-qubit gates. The purpose of the $Z_\frac{\pi}{4}$ operations performed by the phase shifter is to rotate the basis in which the $R_y (\theta)$ gate is applied. Two passes of a photon through the phase shifter corresponds to a rotation on the Bloch sphere (see Figure \ref{fig:ring_computer}(c)) about $\hat z$ by $90^\circ$; this change of basis causes a subsequent $R_y (\theta)$ to effectively rotate about $\hat x$. An additional two passes through the phase shifter rotates $\hat x$ to $- \hat y$, allowing $R_y(\theta)$ to act about $\hat y$ again. The goal here is to construct an operation that has the form $U = R_y (\theta_3) R_x (\theta_2) R_y (\theta_1)$, which is sufficient to implement any single-qubit gate up to an overall phase decomposed via Euler angles~\cite{Jozsa2005AnComputation}.

Consider a sequence of three teleported rotation gates (Eq. \ref{eq:psi_out}) about angles $\theta_1, \theta_2, \theta_3$ which yield measurement results $m_1, m_2, m_3$. As we build up the target operator $U$ with these successive rotations, the outcomes $m_1, m_2, m_3$ can result in extraneous Pauli gates between rotations which effectively offset the target angles $\theta_1, \theta_2, \theta_3$ by $\pi$, as in the second line of Eq. \ref{eq:psi_out}. Intuitively, this is equivalent to constructing an arbitrary rotation in 3D space using only fixed $90^\circ$ rotations about $\hat z$, together with variable rotations about $\hat y$ which may overshoot by $\pi$.

Borrowing a concept from measurement-based quantum computation \cite{Jozsa2005AnComputation, Nielsen2006Cluster-stateComputation, raussendorf_one-way_2001}, we apply rotations to the atomic qubit about \emph{adaptive angles} of $\theta_2 \left(m_1\right)$ and $\theta_3 \left(m_2, m_1\right)$, each of which depends on the results of the preceding measurements. This allows us to propagate the Pauli errors from the middle of the gate to the front and consolidate them as a single error term. The sequence of three rotations performed in this adaptive basis thus implements the operation:
\begin{align}
\begin{split}
\label{eq:three_rotations}
U = \;& \varepsilon(m_3, m_2, m_1) \times \\
& Z_{\frac{\pi}{4}} R_y\left(\theta_3(m_2, m_1)\right) R_x\left(\theta_2(m_1)\right) R_y(\theta_1) Z_{\frac{\pi}{4}},
\end{split}
\end{align}
where the rotations are implicitly programmed to implement $U$ in the basis rotated by $Z_{\frac{\pi}{4}}$ and where the error term $\varepsilon(m_3, m_2, m_1)$ is $\sigma_x$, $\sigma_y$, or $\sigma_z$ up to a global phase. This error term $\varepsilon$ can then either (i) be implicitly removed by programming a subsequent gate $U^\prime$ to instead implement $U^\prime \varepsilon^{-1}$ or (ii) be explicitly removed by scattering the photon against the atom initialized in the non-interacting $\ket{g_0}$ state or in the fully-interacting $\ket{g_1}$ state, applying $\sigma_x$ or $\sigma_z$, respectively. The full derivation for this gate construction process is shown in much greater detail in the Supplementary Information.~\cite{SM}


\subsection{Two-photon gates}
In addition to implementing single-qubit gates, a two-photon entangling gate is needed for universal computation. A controlled phase-flip gate $\cz$ between two photonic qubits $j$ and $k$ can be constructed through a sequence of three scattering interactions in a manner similar to the protocol described by Duan and Kimble \cite{Duan2004ScalableInteractions}. However, the beamsplitter and phase shifter, which are needed to implement the single-qubit gates in our scheme, only allow us to apply operations of the form shown in Eq. \ref{eq:phi_pre_measurement} to the photon-atom system with each pass of a photon through the scattering unit. This prevents us from performing the exact protocol described in Ref. \onlinecite{Duan2004ScalableInteractions} despite the similarities of the proposed physical systems.

We can resolve this complication by modifying the protocol to terminate with a measurement on the atom. We denote the operation applied to the photon-atom state by a pass of photon $j$ through the scattering unit interacting with the atom $a$ as:
\begin{equation}
\label{eq:zeta}
\zeta^{ja} \equiv \left( Z_\frac{\pi}{4} B \right)^j \cz^{ja} \left( B Z_\frac{\pi}{4} \right)^j.
\end{equation}
To implement $\cz^{jk}$ between photons $j$ and $k$, we pass photon $j$ through the scattering unit, then $k$, then $j$ again, separated by $R_y (\pm \frac{\pi}{2})$ rotations applied to the atom. \cite{SM} This results in the state
\begin{equation}
\zeta^{ja} R_y^a (\pi/2) \zeta^{ka} R_y^a (-\pi/2) \zeta^{ja} \left( \ket{\psi_{jk}} \otimes \ket{+}\right),
\end{equation}
where $\ket{\psi_{jk}}$ is the arbitrary state of photons $j$ and $k$ and where the atom is initialized to $\ket{+}$. After this scattering sequence, we measure the state of the atom, which projects the two-photon state to:
\begin{multline}
\left( Z_\frac{\pi}{4} B \otimes Z_\frac{\pi}{4} B \right) \left( B Z_{(-1)^m \frac{\pi}{2}} B \otimes \id \right) \times \\ 
\cz^{jk} \times \left( B Z_\frac{\pi}{4} \otimes B Z_\frac{\pi}{4} \right) \ket{\psi_{jk}},
\label{eq:cz}
\end{multline}
where the extraneous single qubit terms $B Z_\frac{\pi}{4}$, $Z_\frac{\pi}{4} B$, and $B Z_{(-1)^m \frac{\pi}{2}} B$ are artifacts of the photons passing through the beamsplitter and phase shifter. These extra gates are not problematic: when constructing a circuit from single-qubit gates and $\cz$, they may be removed by programming previous and subsequent single-qubit gates to include the inverse gates.

It is worth noting two alternative implementations of the photon-photon $\cz$ gate. First, using the SWAP operation implemented by Eq. \ref{eq:swap}, the states of one photonic qubit and the atom can be exchanged, and the second photon can directly interact with the state of the first, as discussed in the supplementary information\cite{SM}. Second, the protocol demonstrated by Ref. \onlinecite{hacker_photonphoton_2016} can be implemented on this system, reducing the amortized number of scattering passes per $\cz$.

Our proposed device can thus implement arbitrary single-qubit gates and a two-photon $\cz$ gate. This comprises a universal gate set~\cite{Barenco1995ElementaryComputation}, so the device can perform any quantum computation. 

\begin{figure}[t]
\centering
\includegraphics[width=\columnwidth]{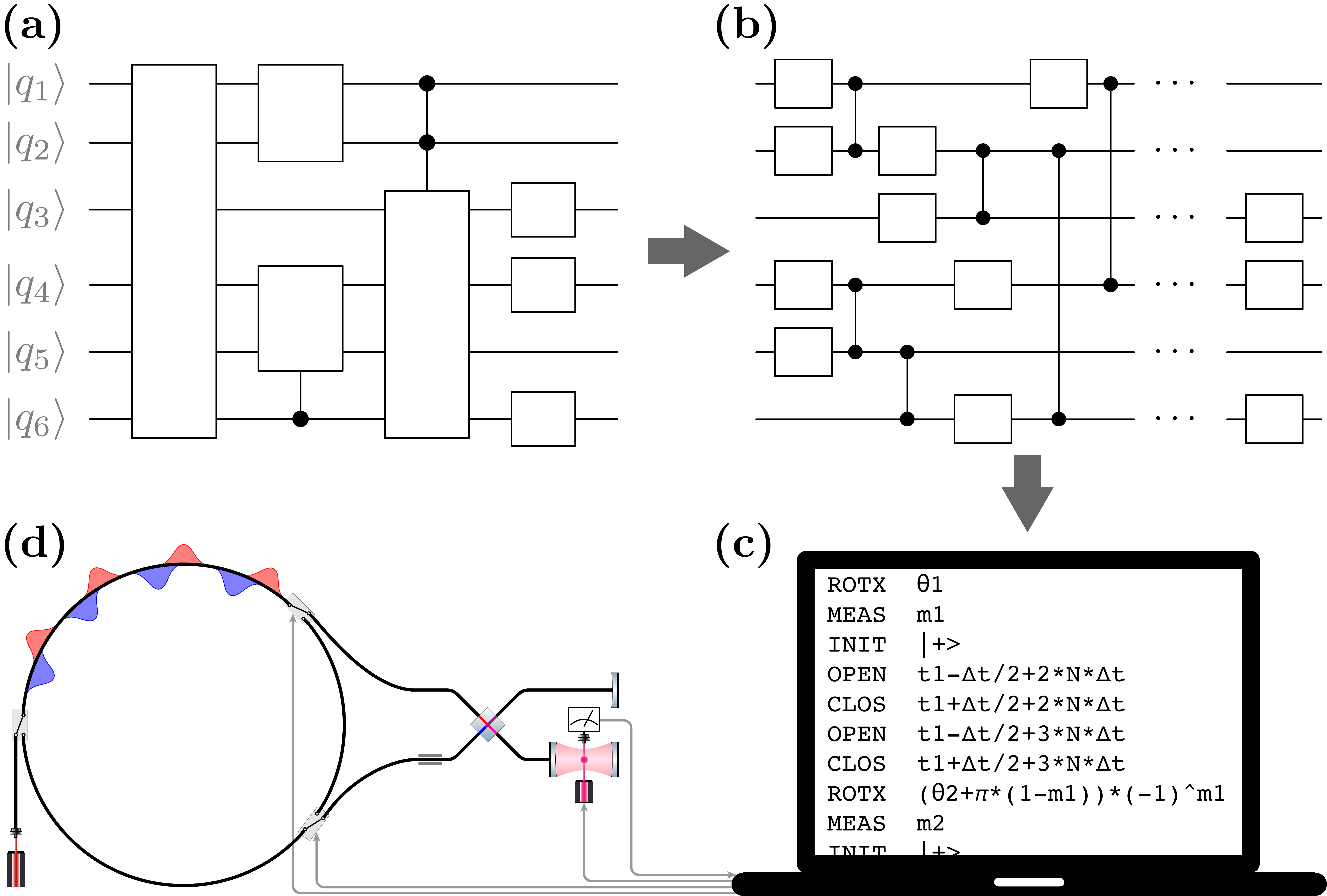}
\caption{
Conceptual illustration of compiling a quantum circuit into an instruction sequence to be performed on the device. 
\textbf{(a)}
A generic target quantum circuit. 
\textbf{(b)}
Decomposition into an equivalent circuit of single-qubit and $\cz$ gates.
\textbf{(c)}
The circuit is further decomposed into a sequence of scattering interactions. This sequence can be assembled on a classical computer into an instruction set with six distinct primitives which correspond to physical actions.
\textbf{(d)}
The controllable elements of the quantum device are the optical switches, cavity laser, and atomic state readout.
}
\label{fig:circuit_compilation}
\end{figure}

\subsection{Arbitrary circuit compilation}
To implement an arbitrary $n$-qubit operator $U \in \mathrm{U}(2^n)$, one could employ a three-step circuit compilation process outlined in Figure \ref{fig:circuit_compilation}. First, decompose $U$ into a sequence of single-qubit gates and $\cz$ operations. This is a well-studied problem~\cite{Mottonen2005DecompositionsGates} and can be done using the same operator preparation routine described in our previous work~\cite{Bartlett2020UniversalProcessing}, but with an additional $\mathcal{O}(n)$ speedup, as this scheme has all-to-all instead of nearest-neighbor connectivity between qubits. Second, represent each $\cz$ as in Eq.~\ref{eq:cz} and decompose each single-qubit gate via Euler angles into rotations which may be teleported onto the photonic qubits. Finally, use a classical control system to modify the adaptive rotations which are applied to the atomic qubit based on the measurement outcomes during operation and to explicitly correct for $\varepsilon$ Pauli errors when necessary. A more detailed discussion of the compilation process and an example instruction sequence to implement a three-qubit quantum Fourier transform can be found in the Supplementary Information.~\cite{SM}

\subsection{Quantum state readout}
After applying the desired quantum operation using the circuit compilation routine outlined above, the state of the photonic qubits must be measured to obtain a classical result. This can be done without the need for single photon detectors with their limited detection efficiencies by sequentially swapping the quantum states of each photonic qubit with that of the atom and repeatedly measuring the atomic state. To perform this SWAP operation, we scatter the desired photonic qubit $j$ against the atom three times; between scattering operations, we apply the rotation $R_y (\pi/2) R_x (\pi)$ to the atomic qubit. Denoting this rotation as $\rho^a$ and using $\zeta^{ja}$ as defined in Eq. \ref{eq:zeta}, it is easily verified that
\begin{equation}
\label{eq:swap}
( B Z_{\frac{\pi}{4}} )^j \zeta^{ja} \rho^a \zeta^{ja} \rho^a \zeta^{ja} (  Z_{\frac{\pi}{4}} B )^j = e^{i \pi}
\left(
\begin{smallmatrix}
1 & 0 & 0 & 0 \\
0 & 0 & 1 & 0 \\
0 & 1 & 0 & 0 \\
0 & 0 & 0 & 1
\end{smallmatrix}
\right),
\end{equation}
which is the SWAP operation up to a factor of -1. Here, $( B Z_{\frac{\pi}{4}} )^j$ and $( Z_{\frac{\pi}{4}} B )^j$ are the operations applied to photon $j$ on the outgoing and return trip from the scattering unit, respectively.\cite{Note3}

\section{Imperfection analysis} 

We now present a theoretical model to analyze the performance of our scheme in the presence of experimental imperfections. The main sources of error for our proposed scheme can be grouped into three classes: (i) deformation of the input pulse shape after scattering off the atom-cavity system, (ii) atomic spontaneous emission loss, and (iii) photon leakage due to attenuation and insertion loss while propagating through the storage ring and optical switches.

\begin{figure}[t!]
\centering
\includegraphics[width=\columnwidth]{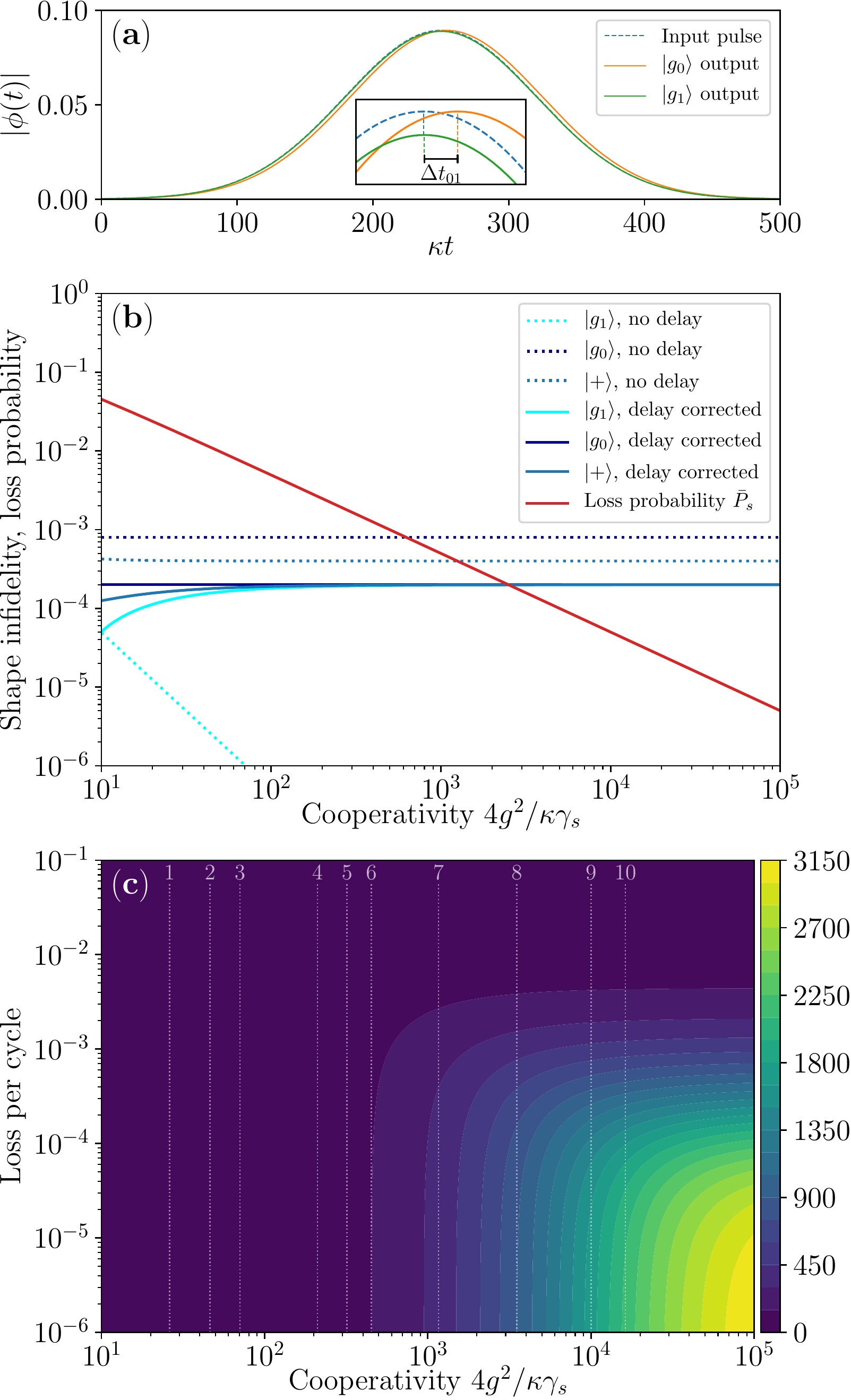}
\caption{
\textbf{(a)}
Output pulse shapes for $\ket{g_0}$ and $\ket{g_1}$ initialization when a cavity with cooperativity $C=180$ is driven by a Gaussian input pulse. The inset highlights the behavior near maximum: the $\ket{g_0}$ output pulse is delayed and the $\ket{g_1}$ output has reduced amplitude.
\textbf{(b)}
Shape infidelity and photon leakage probability as a function of cavity cooperativity. Solid blue lines show the pulse shape infidelity when the reference pulse is delayed by $\Delta t_{01} / 2$.
\textbf{(c)}
Estimated single-qubit circuit depth achievable while maintaining $>50\%$ fidelity as a function of cavity cooperativity and photon attenuation per cycle, assuming one scattering interaction every cycle and no error correction. Dotted lines show various experimentally demonstrated cooperativity values in similar cavity systems. Lines 1-10 correspond respectively to Refs. \onlinecite{hacker_photonphoton_2016}, \onlinecite{covey_telecom-band_2019}, \onlinecite{samutpraphoot_strong_2020}, \onlinecite{mckeever_experimental_2003}, \onlinecite{fushman_controlled_2008}, \onlinecite{miller_trapped_2005}, \onlinecite{colombe_strong_2007}, \onlinecite{li_vacuum_2018}, \onlinecite{chang_colloquium_2018}, and \onlinecite{suleymanzade_tunable_2020}.
}
\label{fig:error_analysis}
\end{figure}

In our analysis, we assume the cavity mode frequency $\omega_c$ is exactly resonant with the atom $\ket{g_1} \leftrightarrow \ket{e}$ transition frequency $\Omega_1$, since the detuning can be calibrated to be zero in both free-space and nanophotonic systems~\cite{samutpraphoot_strong_2020}. We also assume that rotations of the atomic state using the cavity laser and measurement of the state can be done with fidelity $\mathcal{F} \approx 1$, as both processes have been demonstrated experimentally with infidelities significantly lower than the error sources listed above~\cite{Bruzewicz2019Trapped-ionChallenges,gehr_cavity-based_2010,Bermudez2017AssessingComputation,Myerson2008High-fidelityQubits,Harty2014High-fidelityBit,Burrell2010ScalableFidelity}. For all simulations here, we choose a photon pulse width of $\tau = 100 / \kappa$, a time range $T = 500/\kappa$, and compute cooperativity with fixed $\gamma_s = \kappa/5$. This choice of parameters were motivated by a sample of experimental cavity setups enumerated in Figure \ref{fig:error_analysis} and result in a temporal bin size of order $100$ nanoseconds for $\kappa / 2\pi \sim 1 \text{GHz}$. Greater detail is given in the Supplementary Information~\cite{SM}.

We use the analytical technique described by Shen and Fan~\cite{Shen2009TheoryAtom, Shen2009TheoryAtomb} to exactly solve the single-photon transport problem of the coupled atom-cavity-waveguide system and obtain the output pulse $\phi_\text{out} (t)$ when the system is driven by an input pulse $\phi_\text{in} (t)$. This treatment captures the full quantum mechanical response of the system to a single-photon input Fock state for an arbitrary initialization of the atom without making the semiclassical assumption of a weak coherent input state.

Figure \ref{fig:error_analysis}(a) shows the output pulse shapes for a single-photon Gaussian input pulse when the atom is initialized in states $\ket{g_0}$ or $\ket{g_1}$. For the $\ket{g_0}$ initialization, the response is identical to an empty cavity, since the $\ket{g_0} \leftrightarrow \ket{e}$ transition is far-detuned from the cavity resonant frequency. In this case, the output pulse is slightly delayed from the input pulse. For the $\ket{g_1}$ initialization, the photon is directly reflected from the front mirror of the cavity since the dressed cavity modes are well-separated in the strong coupling limit from the input photon frequency by the vacuum Rabi splitting, so the delay is minimal. We denote the difference in the delays of the $\ket{g_0}$ and $\ket{g_1}$ scatterings as $\Delta t_{01}$ (see inset). We compute the pulse shape fidelity $\mathcal{F}_\text{shape}$ as the overlap integral of the output pulse with the input pulse after both pulses have been normalized to have unit area, and the pulse shape infidelity is $1-\mathcal{F}_\text{shape}$. This quantity only describes mismatch of the shapes of the input and output pulses, not mismatch of the pulse areas; the infidelity due to photon loss is computed as a separate quantity.

In Figure \ref{fig:error_analysis}(b), we plot the shape infidelity of various states as a function of the single-atom cavity cooperativity $C = 4g^2 / \kappa \gamma_s$, where $\kappa$ is the decay rate of the cavity into the waveguide, $g$ is the atom-cavity coupling strength, and $\gamma_s$ is the atomic spontaneous emission rate. The pulse shape infidelity from scattering off the $\ket{g_1}$ state decreases to negligible values as $C$ increases, while the infidelity of $\ket{g_0}$ reaches an asymptote at $8\times 10^{-4}$ due to the delay of the output pulse by a time which is independent of $C$. The infidelity from scattering against $\ket{+} = (\ket{g_0}+\ket{g_1}) / \sqrt{2}$ thus reaches a value of $4 \times 10^{-4}$. Since the atom will usually be initialized to the $\ket{+}$ state during operation of the device, it is desirable to minimize the infidelity of this interaction. This can be done by equally distributing the delays between the $\ket{g_0}$ and $\ket{g_1}$ states by delaying the reference pulse by a time difference $\Delta t_{01} / 2$, adding path length $c \, \Delta t_{01} / 4$ to the top waveguide in Figure \ref{fig:ring_computer}(a). This results in a ``delay corrected'' infidelity of $2 \times 10^{-4}$, which is independent of both cavity cooperativity (for $C \gg 1$) and atomic state initialization.

In Figure \ref{fig:error_analysis}(b), we also plot the photon leakage probability for a scattering interaction. Atomic spontaneous emission noise from the excited $\ket{e}$ state into modes other than the cavity mode at a rate $\gamma_s$ results in a partial loss of the photon, resulting in an output pulse with total photon number $\int dt\, |\phi_{\rm out}(t)|^2 < 1$. We calculate the probability $P_s$ of spontaneous emission loss as $ P_s = 1 - \frac{\int dt\, |\phi_{\rm out}(t) |^2 }{\int dt\, |\phi_{\rm in}(t)|^2}$. Spontaneous emission noise only applies to the $\ket{1} \otimes \ket{g_1}$ component of the $\text{photon} \otimes \text{atom}$ state; since the atom will usually be initialized to the $\ket{+}$ state, if we average over the possible input photon states, we obtain an average leakage probability of $\bar P_s = P_s/4$. This average photon loss probability is plotted as the red line in Figure \ref{fig:error_analysis}(b) and ranges from about $5\%$ to $0.0005\%$ over the range of cooperativity values shown.

Finally, we account for loss due to attenuation in the optical storage ring and insertion loss from the switches as an average loss per cycle $L$. To estimate the maximum circuit depth $D$ attainable with an overall fidelity $\mathcal{F}_\text{target}$, we compute a ``bulk fidelity'' per cycle accounting for shape infidelity, spontaneous emission loss, and attenuation while propagating through the storage ring and optical switches. For simplicity, we assume the circuit operates on only a single photonic qubit and that the photon is scattered against the atom with every pass through the storage ring. The achievable circuit depth operating with success probability $\mathcal{F}_\text{target}$ is thus the maximum $D$ satisfying $\left [\mathcal{F}_\text{shape} \times (1 - \bar P_s) \times (1 - L) \right]^{D} \ge \mathcal{F}_\text{target}$, which is plotted as a function of cavity cooperativity and propagation loss in Figure \ref{fig:error_analysis}(c) for $\mathcal{F}_\text{target} = 50\%$. Using optimistic but not unrealistic values for cooperativity~\cite{chang_microring_2019, chang_colloquium_2018, suleymanzade_tunable_2020} $C=10^4$ and cycle loss $L = 10^{-4}$, we compute a bulk fidelity of $\mathcal{F} \approx 99.95\%$. This allows for an estimated depth of $D \approx 2000$ scattering operations while maintaining $50\%$ success rate, and results in an error probability per gate (EPG) of $\sim 5 \times 10^{-4}$, below the estimated $\sim 10^{-3}$ EPG threshold for fault tolerance~\cite{Nielsen2010QuantumInformation,Gottesman1997StabilizerCorrection, Preskill1997Fault-tolerantComputation, Knill2005QuantumDevices}. Additionally, photon loss, which is likely the main error mechanism, can be efficiently corrected up to a per-gate loss of $\sim10^{-2}$ using concatenated codes~\cite{Knill2000ThresholdsComputation}.

More broadly, generalizations of the proposed scheme to synthetic dimensions other than time multiplexing could further improve the scalability of our design. Instead of using counter-propagating optical modes, one could encode each qubit in the polarization basis. With suitable design of the atom-cavity interacting system, frequency~\cite{yuan_photonic_2016, ozawa_synthetic_2016} or angular momentum modes~\cite{luo_quantum_2015} could be used as an alternative synthetic dimension. These concepts would naturally lend themselves to studying quantum many-body physics of interacting Hamiltonians in synthetic space~\cite{ozawa_topological_2019}, which is difficult to realize in purely photonic platforms without the strong single-photon nonlinearity of the atom which we employ here~\cite{yuan_creating_2019, ozawa_synthetic_2017}.

\section{Conclusion}
In this paper we have presented a scheme for universal quantum computation using a single coherently controlled atom to indirectly manipulate a many-photon quantum state. We have shown that arbitrary single-qubit gates can be deterministically constructed from rotations applied to the atomic qubit and teleported onto the photonic qubits via projective measurements. Using similar scattering processes, two-photon $\cz$ gates can be implemented, and readout of the photonic quantum state can be done using only atomic measurements with efficiencies far greater than that of state-of-the-art photon detectors. Our proposed scheme has high fidelity even in the presence of realistic experimental imperfections and offers significant advantages in required physical resources and experimental feasibility over many existing paradigms for photonic quantum computing.

\vspace{5mm} 

\emph{Acknowledgements} ---
This work was supported by a Vannevar Bush Faculty Fellowship from the U.S. Department of Defense (N00014-17-1-3030) and by a U.S. Air Force Office of Scientific Research MURI project (FA9550-17-1-0002).

\emph{Author contributions} ---
B.B. and S.F. conceived the project. B.B. developed the device design and gate mechanisms. A.D. performed the imperfection analysis with contributions from B.B. All authors contributed to writing the manuscript. S.F. supervised the project.

\emph{Competing interests} ---
The authors declare no competing interests.




\end{document}


\title{Supplementary Information for \\
``Deterministic photonic quantum computation in a synthetic time dimension''}

\author{Ben Bartlett}
\email{benbartlett@stanford.edu}
\affiliation{Department of Applied Physics, Stanford University, Stanford, CA 94305, USA}
\affiliation{Department of Electrical Engineering, Stanford University, Stanford, CA 94305, USA}

\author{Avik Dutt}
\affiliation{Department of Electrical Engineering, Stanford University, Stanford, CA 94305, USA}

\author{Shanhui Fan}
\email{shanhui@stanford.edu}
\affiliation{Department of Electrical Engineering, Stanford University, Stanford, CA 94305, USA}

\maketitle

In this Supplementary Information document, we give more detailed presentations of the results described in the main paper. In Section \ref{sec:gate_teleportation_derivation} we present a derivation of the gate teleportation mechanism; in Section \ref{sec:single_qubit_gates} we derive a method to construct arbitrary single-qubit operations from the teleported gates; in Section \ref{sec:swap_gate} we construct a photon-atom SWAP operation from scattering sequences and measurement; in Section \ref{sec:control_z} we describe constructions for a two-photon $\cz$ gate; in Section \ref{sec:circuit_compilation} we give more detail of the circuit compilation process and provide an example of a compiled instruction sequence to implement a quantum Fourier transform on our proposed device; and in Section \ref{sec:imperfection_analysis} we discuss in greater detail the imperfection analysis described in the main text.

\begin{figure*}[h]
\centering 
\includegraphics[width=\textwidth]{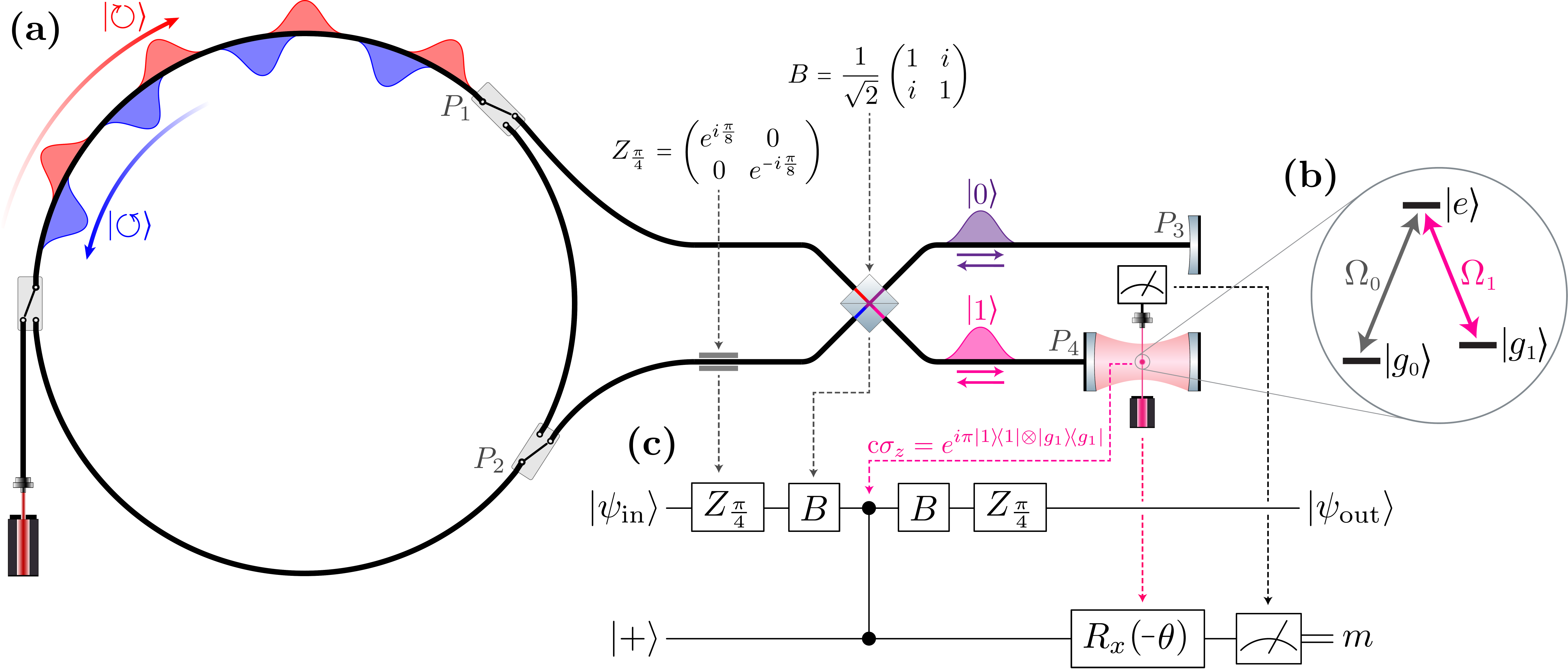}
\caption{
An annotated figure depicting the architecture described in the main text and the correspondence of physical and logical circuit elements.
\textbf{(a)} 
The physical design of the device, with annotations indicating quantum operations implemented by physical circuit elements.
\textbf{(b)} 
The energy structure of the atom: $\Omega_1$ is resonant with the cavity mode and photon carrier frequency, while $\Omega_0$ is far-detuned.
\textbf{(c)} 
Gate diagram of the quantum circuit applied in a single pass of a photonic qubit through the scattering unit. The top rail denotes the state of the photonic qubit and the bottom rail denotes the atomic qubit. After the photon returns to the storage ring, $R_x (-\theta)$ is applied to the atomic qubit and a projective measurement of the atomic state is performed. The final output state $\ket{\psi_\text{out}}$ is $Z_{\frac{\pi}{4}} \sigma_z \left(-\sigma_y \right)^{m\oplus 1} R_y\left(\theta\right) Z_{\frac{\pi}{4}} \ket{\psi_\text{in}}$, as described in Eq. 2 of the main text.
}
\label{fig:ring_computer_SM}
\end{figure*}

\section{Derivation of gate teleportation mechanism}
\label{sec:gate_teleportation_derivation}

Consider a photon which is circulating in the storage ring in the storage ring in a state $\ket{\psi_\text{in}} = \alpha \ket{\CW} + \beta \ket{\CCW}$, where $\ket{\CW}$ and $\ket{\CCW}$ denote the two counter-circulating states. Referring to Figure \ref{fig:ring_computer_SM}, define bosonic operators $\adag_\CW (t)$, $\adag_\CCW (t)$ which create at time $t$ a clockwise- or counterclockwise-propagating photon in the ring at the points $P_1$, $P_2$, respectively, just before the switches. The physical state of the photon in the ring can be written as 
\begin{equation}
\label{eq:photon_physical_input_state}
\ket{\psi_\text{in}} = \int dt \, \phi(t) \left[ \alpha \, \adag_\CW (t) + \beta \, \adag_\CCW (t) \right] \vac,
\end{equation}
where $\vac$ denotes the vacuum state and $\phi(t)$ describes the pulse envelope. Here we assume that the photon was originally injected in the $\ket{\CW}$ state as shown in Figure \ref{fig:ring_computer_SM} and has undergone at most a small number of scattering interactions with the atom-cavity system, such that the clockwise and counterclockwise pulses have not independently deformed significantly and can be described by a single envelope. 

We also define bosonic operators $\bdag_{0,d} (t)$, $\bdag_{1,d} (t)$ which respectively create a photon in the top or bottom waveguides at points $P_1, P_2$ at time $t$ propagating with direction $d \in \{L,R\}$. As the photon is injected by the switches from the ring into the waveguides, the fixed $\pi/4$ phase shifter applies (up to a global phase) a rotation $Z_\frac{\pi}{4} \equiv R_z \left(\frac{\pi}{4}\right) = \smat{e^{-i\pi/8}}{0}{0}{e^{i\pi/8}}$ to the photon state, and the beamsplitter applies the operation $B = \frac{1}{\sqrt{2}} \smat{1}{i}{i}{1}$. Finally, let operators $\cdag_{0,d} (t)$, $\cdag_{1,d} (t)$ with $d \in \{L,R\}$ create a photon at time $t$ in the top or bottom waveguides just before the mirror or cavity at points $P_3$ and $P_4$. 

The round trip distance from points $P_1, P_2$ to $P_3, P_4$ and back is equal to the ring circumference $L = n \Delta t$, where the speed of light in the waveguides is set to unity and where $n$ is the number of time bins. This matching path length ensures that a photon which leaves the ring to scatter against the atom will return to its original time bin. Let time $t_0$ denote the point at which the clockwise and counterclockwise components of the photon in the ring pass their respective switches and may be injected into the scattering unit. When the switches are set to the open state, we can relate the $\adag, \bdag, \cdag$ operators on the outgoing pass of the photon with:
\begin{equation}
\label{eq:operator_couplings_outgoing}
\begin{bmatrix} 
\bdag_{0,R}(t_0) \\ 
\bdag_{1,R}(t_0) 
\end{bmatrix} = \begin{bmatrix} 
\adag_\CW (t) \\ 
\adag_\CCW (t) 
\end{bmatrix}, 
\quad 
\begin{bmatrix} 
\cdag_{0,R} \left(t_0 + \frac{n \Delta t}{2}\right) \\ 
\cdag_{1,R} \left(t_0 + \frac{n \Delta t}{2}\right) 
\end{bmatrix} = B Z_\frac{\pi}{4} \begin{bmatrix} 
\bdag_{0,R}(t) \\ 
\bdag_{1,R}(t) 
\end{bmatrix}.
\end{equation}

The $\cdag_{1,R}$ component of the photon interacts at time $t_1 = t_0 + \frac{n \Delta t}{2}$ with the $\ket{g_1}$ component of the atomic state that is resonant with the photon frequency, implementing the unitary transformation $\cz = e^{i \pi \ketbra{1}{1} \otimes \ketbra{g_1}{g_1}} = \exp\left(i \pi \, \cdag_{1,R} \ket{\emptyset}\bra{\emptyset} \hat{c}_{1,R} \otimes \ketbra{g_1}{g_1}\right)$. Thus, we can relate the operators before and after reflection/scattering as:
\begin{equation}
\label{eq:operator_couplings_reflection}
\left(
\begin{bmatrix} 
\cdag_{0,L} (t_1) \\ 
\cdag_{1,L} (t_1)
\end{bmatrix} 
\otimes 
\begin{bmatrix} 
\ketbra{g_0}{g_0} \\ 
\ketbra{g_1}{g_1}
\end{bmatrix} 
\right)
= \exp\left(i \pi \, \cdag_{1,R}(t_1)\, \hat{c}_{1,R}(t_1) \otimes \ketbra{g_1}{g_1}\right)
\left(
\begin{bmatrix} 
\cdag_{0,R} (t_1) \\ 
\cdag_{1,R} (t_1) 
\end{bmatrix}
\otimes 
\begin{bmatrix} 
\ketbra{g_0}{g_0} \\ 
\ketbra{g_1}{g_1}
\end{bmatrix}
\right),
\end{equation}
where we assume that the interaction timescale (usually set by the cavity lifetime) is negligible compared to the time bin size $\Delta t$ (the long pulse limit). Eq. \ref{eq:operator_couplings_reflection} is derived for scattering in the single-photon subspace, but is applicable to multi-photon states as long as the photon wavefunctions do not overlap in the scattering unit.

On the return trip, after scattering against the atom, the photon passes through the beamsplitter and phase shifter in reverse order before being re-injected at time $t_2 = t_1 + \frac{n \Delta t}{2}$ into the ring at points $P_1, P_2$, allowing us to relate the final set of operators:
\begin{equation}
\label{eq:operator_couplings_return}
\begin{bmatrix} 
\bdag_{0,L}(t_2) \\ 
\bdag_{1,L}(t_2)
\end{bmatrix} = Z_\frac{\pi}{4}^\transpose B^\transpose \begin{bmatrix} 
\cdag_{0,L} (t_1) \\ 
\cdag_{1,L} (t_1)
\end{bmatrix}, 
\quad 
\begin{bmatrix} 
\adag_\CCW (t_2) \\ 
\adag_\CW (t_2)
\end{bmatrix} = \begin{bmatrix} 
\bdag_{0,L}(t_2) \\ 
\bdag_{1,L}(t_2)
\end{bmatrix}.
\end{equation}

Note that the $\adag$ and $\bdag$ operators have opposite couplings on the photon's return trip; e.g. the clockwise $\adag_\CW$ operator couples to the top waveguide $\bdag_{0,R}$ on the outgoing direction, while on the return trip, the top waveguide $\bdag_{0,L}$ couples to the counterclockwise mode $\adag_\CCW$. One can combine the equations above to obtain that, if the atom is in the non-interacting state $\ket{g_0}$, the total transformation performed on the photon by a round trip through the scattering unit is $Z_\frac{\pi}{4} B B Z_\frac{\pi}{4} $, and the photon state in the ring is unchanged up to a factor of $i$: $\adag_\CW (t + n \Delta t) = i \adag_\CW (t)$ and $\adag_\CCW (t + n \Delta t) = i \adag_\CCW (t)$.

For the purpose of the gate teleportation, we initialize the atom in the $\ket{g_0}$ state and use a $R_y \left(\pi/2\right)$ rotation to change the state to $\ket{+} \equiv \frac{1}{\sqrt{2}} \left(\ket{g_0} + \ket{g_1}\right)$. The scattering interaction applies a $\pi$ phase shift to the $\ket{1}\otimes\ket{g_1}$ component of the joint quantum state, implementing a $\cz$ gate. After the photon has interacted with the atom, an $R_x(-\theta)$ rotation is applied to the atom as the photon passes back through the beamsplitter and phase shifter and is injected back into the ring.  Thus, the joint photon-atom state after scattering is:
\begin{equation}
\label{eq:state_pre_measurement}
\ket{\Phi} = \left( (Z_\frac{\pi}{4} B) \otimes R_x(-\theta) \right) \cz \left( (B Z_\frac{\pi}{4}) \otimes R_y(\pi/2) \right) \left(\ket{\psi_\text{in}} \otimes \ket{g_0} \right).
\end{equation}

Finally, a projective measurement of the atom's state in the $\left\{\ket{g_0}, \ket{g_1}\right\}$ basis is performed, obtaining a bit $m\in\{0,1\}$. If the atomic state collapses to state $\ket{g_m}$, then we obtain a disentangled output photon-atom state:
\begin{equation}
\ket{\psi_\text{out}} \otimes \ket{g_m} = \frac{1}{\sqrt{P_m}} \left[\id \otimes \ketbra{g_m}{g_m} \right] \ket{\Phi},
\end{equation}
where $P_m = \tr \left[(\id \otimes \ketbra{g_m}{g_m}) \ketbra{\Phi}{\Phi} \right]$. Working in the long pulse, high cooperativity limit where pulse shape deformation from the scattering interaction is negligible\footnote{
Here we assume that the temporal pulse length $\tau$ is much less than the time bin spacing $\Delta t$ but significantly larger than the cavity decay rate, such that the pulse shape changes slowly compared to the cavity decay rate. This means that the pulse shapes for the clockwise and counterclockwise components of the photon state do not change independently. A more realistic treatment of the pulse deformation is given in the imperfection analysis presented here and in the main text.
}, we obtain respective output states for $m = 0,1$ of:
\begin{align}
\label{eq:psi_out_full}
\ket{\psi_\text{out}} \otimes \ket{g_0} &= \int dt \; \phi(t) \left[
\left(i \beta \cos \frac{\theta}{2}  + e^{\frac{i \pi}{4}} \alpha \sin \frac{\theta}{2}\right) \adag_\CW (t) + 
\left(i \alpha \cos \frac{\theta}{2} + e^{-\frac{i \pi}{4}} \beta \sin \frac{\theta}{2}\right) \adag_\CCW (t)
\right] \vac \otimes \ket{g_0} \\ 
\ket{\psi_\text{out}} \otimes \ket{g_1} &= \int dt \; \phi(t) \left[
\left(e^{-\frac{i \pi}{4}} \alpha \cos \frac{\theta}{2} - \beta \sin \frac{\theta}{2}\right) \adag_\CW (t)  
-\left(e^{ \frac{i \pi}{4}} \beta \cos \frac{\theta}{2} + \alpha \sin \frac{\theta}{2}\right) \adag_\CCW (t) 
\right] \vac \otimes \ket{g_1},
\end{align}
with $\alpha, \beta$ the coefficients from the input state of Eq. \ref{eq:photon_physical_input_state}. Thus, the output photon state $\ket{\psi_\text{out}}$, depending on the outcome of the atomic measurement $m$, is:
\begin{align}
\begin{split}
\label{eq:psi_out}
\ket{\psi_\text{out}} &= 
\begin{cases}
-i Z_{\frac{\pi}{4}} \sigma_z R_y(\theta+\pi) Z_{\frac{\pi}{4}} \ket{\psi_\text{in}} & \text{ if } m=0 \\ 
Z_{\frac{\pi}{4}} \sigma_z R_y(\theta) Z_{\frac{\pi}{4}} \ket{\psi_\text{in}} & \text{ if } m=1
\end{cases} \\
&= Z_{\frac{\pi}{4}} \sigma_z \left(-\sigma_y \right)^{m\oplus 1} R_y\left(\theta\right) Z_{\frac{\pi}{4}} \ket{\psi_\text{in}},
\end{split}
\end{align}
where $m\oplus 1$ denotes addition modulo 2.

\section{Constructing arbitrary single-qubit rotations}
\label{sec:single_qubit_gates}

To construct arbitrary single-qubit gates, we compose a sequence of teleported gates of the form in Eq. \ref{eq:psi_out} with a sequence of ``non-entangling'' scattering process which correct for local Pauli errors introduced depending on the atomic measurement outcomes. If the atom is initialized to the off-resonant $\ket{g_0}$ state, then the atom-cavity system is on resonance with the incident photon and behaves as a mirror. In this case, the $\pi$ phase shifts imparted by the cavity and by the mirror in the top waveguide cancel, and the photon state is transformed as $\ket{\psi_\text{out}} = Z_\frac{\pi}{4} B B Z_\frac{\pi}{4} \ket{\psi_\text{in}} = i \sigma_x \ket{\psi_\text{in}}$. If the atom is initialized to $\ket{g_1}$, then the atom-cavity system is off resonance with the incident photon. In this case, the phase shift from the mirror in the top waveguide is not matched and a relative $\pi$ phase shift is imparted between the top and bottom modes, transforming the photon state as $\ket{\psi_\text{out}} = Z_\frac{\pi}{4} B \sigma_z B Z_\frac{\pi}{4} \ket{\psi_\text{in}} = -i \sigma_z Z_{\pi/2} \ket{\psi_\text{in}}$.

Now consider a sequence of three successive teleported rotation gates $R_y(\theta_1), R_y(\theta_2), R_y(\theta_3)$, with atomic measurement results $m_1, m_2, m_3$. The goal here is to create a sequence of scattering operations which result in a gate of the form $U = R_y (\theta_3) R_x (\theta_2) R_y (\theta_1)$, which is sufficient to implement any single-qubit gate up to an overall phase decomposed as Euler angles.~\cite{Jozsa2005AnComputation} 
The total operation $U$ applied to the initial input state $\ket{\psi_\text{in}}$ from the three scattering operations is:
\begin{equation}
\label{eq:gate_sequence}
U = 
\left(-1\right)^{m_1\oplus m_2 \oplus m_3 \oplus 1}
Z_{\frac{\pi}{4}} \sigma_z (\sigma_y)^{m_3 \oplus 1} R_y(\theta_3) Z_{\frac{\pi}{4}} 
Z_{\frac{\pi}{4}} \sigma_z (\sigma_y)^{m_2 \oplus 1} R_y(\theta_2) Z_{\frac{\pi}{4}} 
Z_{\frac{\pi}{4}} \sigma_z (\sigma_y)^{m_1 \oplus 1} R_y(\theta_1) Z_{\frac{\pi}{4}} .
\end{equation}
We can simplify this expression by noting that $Z_{\frac{\pi}{4}} Z_{\frac{\pi}{4}} \sigma_z (\sigma_y)^{m \oplus 1} = -i (-i \sigma_y \sigma_z)^{m\oplus 1} Z_{-\frac{\pi}{2}} = -i (\sigma_x)^{m \oplus 1} Z_{-\frac{\pi}{2}} $, which reduces Eq. \ref{eq:gate_sequence} to:
\begin{equation}
U = 
\left(-1\right)^{m_3\oplus m_2 \oplus m_1} \left(-i\right)^{m_2\oplus m_1}
Z_{\frac{\pi}{4}} \sigma_z (\sigma_y)^{m_3 \oplus 1} R_y(\theta_3) 
(\sigma_y \sigma_z)^{m_2 \oplus 1} Z_{-\frac{\pi}{2}}
R_y(\theta_2) 
(\sigma_y \sigma_z)^{m_1 \oplus 1} Z_{-\frac{\pi}{2}}
R_y(\theta_1) Z_{\frac{\pi}{4}} .
\end{equation}

Since the results of previous measurements can add extraneous Pauli gates which affect future rotations, we wish to perform adaptive operations based on the measured outcomes. After the first measurement $m_1$ is performed, the gate operation is:
\begin{equation}
U = 
\begin{cases}
\left(-1\right)^{m_3\oplus m_2} \left(-i\right)^{m_2}
Z_{\frac{\pi}{4}} \sigma_z (\sigma_y)^{m_3 \oplus 1} R_y(\theta_3) 
(\sigma_y \sigma_z)^{m_2 \oplus 1} Z_{-\frac{\pi}{2}}
R_y(\theta_2) 
\sigma_y \sigma_z Z_{-\frac{\pi}{2}}
R_y(\theta_1) Z_{\frac{\pi}{4}} & \text{ if } m_1=0 
\\ 
\left(-1\right)^{m_3\oplus m_2 \oplus 1} \left(-i\right)^{m_2 \oplus 1}
Z_{\frac{\pi}{4}} \sigma_z (\sigma_y)^{m_3 \oplus 1} R_y(\theta_3) 
(\sigma_y \sigma_z)^{m_2 \oplus 1} Z_{-\frac{\pi}{2}}
R_y(\theta_2) 
Z_{-\frac{\pi}{2}}
R_y(\theta_1) Z_{\frac{\pi}{4}} & \text{ if } m_1=1 .
\end{cases}
\end{equation}
Using the identities that $\sigma_z Z_{-\frac{\pi}{2}} = i Z_{+\frac{\pi}{2}}$ and that $R_i (\theta) \sigma_i = -i R_i (\theta+\pi)$ for $i=x,y,z$, we can rewrite this as:
\begin{equation}
U = 
\begin{cases}
\left(-1\right)^{m_3\oplus m_2} \left(-i\right)^{m_2}
Z_{\frac{\pi}{4}} \sigma_z (\sigma_y)^{m_3 \oplus 1} R_y(\theta_3) 
(\sigma_y \sigma_z)^{m_2 \oplus 1} Z_{-\frac{\pi}{2}}
R_y(\theta_2+\pi) 
Z_{+\frac{\pi}{2}}
R_y(\theta_1) Z_{\frac{\pi}{4}} & \text{ if } m_1=0 
\\ 
\left(-1\right)^{m_3\oplus m_2 \oplus 1} \left(-i\right)^{m_2 \oplus 1}
Z_{\frac{\pi}{4}} \sigma_z (\sigma_y)^{m_3 \oplus 1} R_y(\theta_3) 
(\sigma_y \sigma_z)^{m_2 \oplus 1} Z_{-\frac{\pi}{2}}
R_y(\theta_2) 
Z_{-\frac{\pi}{2}}
R_y(\theta_1) Z_{\frac{\pi}{4}} & \text{ if } m_1=1 .
\end{cases}
\end{equation}
Substituting $Z_{-\frac{\pi}{2}} R_y (\theta) Z_{+\frac{\pi}{2}} = R_x (\theta)$ and $Z_{-\frac{\pi}{2}} R_y (\theta) Z_{-\frac{\pi}{2}} = i \sigma_z R_x (-\theta)$, we rearrange the equation to turn the second rotation gate into a $R_x (\pm \theta)$ gate, where the sign depends on the outcome of $m_1$, which is already known:
\begin{align}
\begin{split}
\label{eq:cases1}
U &= 
\begin{cases}
\left(-1\right)^{m_3\oplus m_2} \left(-i\right)^{m_2}
Z_{\frac{\pi}{4}} \sigma_z (\sigma_y)^{m_3 \oplus 1} R_y(\theta_3) 
(\sigma_y \sigma_z)^{m_2 \oplus 1} 
R_x(\theta_2+\pi) 
R_y(\theta_1) Z_{\frac{\pi}{4}} & \text{ if } m_1=0 
\\ 
i \left(-1\right)^{m_3\oplus m_2 \oplus 1} \left(-i\right)^{m_2 \oplus 1}
Z_{\frac{\pi}{4}} \sigma_z (\sigma_y)^{m_3 \oplus 1} R_y(\theta_3) 
(\sigma_y \sigma_z)^{m_2 \oplus 1} 
\sigma_z R_x(-\theta_2) 
R_y(\theta_1) Z_{\frac{\pi}{4}} & \text{ if } m_1=1
\end{cases}
\\
&=
\left(-1\right)^{m_3}
Z_{\frac{\pi}{4}} \sigma_z (\sigma_y)^{m_3 \oplus 1} R_y(\theta_3) 
(i \sigma_y \sigma_z)^{m_2 \oplus 1} \times 
\begin{cases}
R_x(\theta_2+\pi) 
R_y(\theta_1) Z_{\frac{\pi}{4}} & \text{ if } m_1=0 
\\ 
\sigma_z R_x(-\theta_2) 
R_y(\theta_1) Z_{\frac{\pi}{4}} & \text{ if } m_1=1 .
\end{cases}
\end{split}
\end{align}
Importantly, the decision for which adaptive changes to apply to the $\theta_2$ operation (adding $\pi$ or inverting the angle) can be made knowing only the outcome of the previous measurement $m_1$. Let $\theta_2(m_1) = \theta_2 + \pi$ if $m_1=0$ and $-\theta_2$ if $m_1=1$ denote the adaptive angle to implement the desired rotation $R_x(\theta_2)$. Then we can rewrite Eq. \ref{eq:cases1} as:
\begin{equation}
U =
\left(-1\right)^{m_3}
Z_{\frac{\pi}{4}} \sigma_z (\sigma_y)^{m_3 \oplus 1} R_y(\theta_3) 
(i \sigma_y \sigma_z)^{m_2 \oplus 1} \sigma_{z}^{m_1} 
R_x\left(\theta_2(m_1)\right) R_y(\theta_1) Z_{\frac{\pi}{4}}.
\end{equation}

We repeat this process of performing a measurement and commuting the error terms to the front of the equation for measurement $m_2$. After performing the second measurement, we use $R_y (\theta) \sigma_z = \sigma_z R_y (-\theta)$ and the above identities to obtain:
\begin{align}
\begin{split}
\label{eq:cases2}
U &= 
\left(-1\right)^{m_3}
Z_{\frac{\pi}{4}} \sigma_z (\sigma_y)^{m_3 \oplus 1} R_y(\theta_3) \times 
\begin{cases}
i \sigma_y \sigma_z 
R_x\left(\theta_2(m_1)\right) 
R_y(\theta_1) Z_{\frac{\pi}{4}} 
& \text{ if } m_1=0, m_2=0 
\\ 
R_x\left(\theta_2(m_1)\right) 
R_y(\theta_1) Z_{\frac{\pi}{4}} 
& \text{ if } m_1=0, m_2=1 
\\ 
i \sigma_y \sigma_z \sigma_z 
R_x\left(\theta_2(m_1)\right) 
R_y(\theta_1) Z_{\frac{\pi}{4}} 
& \text{ if } m_1=1, m_2=0
\\
\sigma_z
R_x\left(\theta_2(m_1)\right) 
R_y(\theta_1) Z_{\frac{\pi}{4}} 
& \text{ if } m_1=1, m_2=1
\end{cases} 
\\
&= 
\left(-1\right)^{m_3}
Z_{\frac{\pi}{4}} \sigma_z (\sigma_y)^{m_3 \oplus 1} \times 
\begin{cases}
-R_y(\theta_3+\pi)
\sigma_z 
R_x\left(\theta_2(m_1)\right) 
R_y(\theta_1) Z_{\frac{\pi}{4}} 
& \text{ if } m_1=0, m_2=0 
\\ 
R_y(\theta_3)
R_x\left(\theta_2(m_1)\right) 
R_y(\theta_1) Z_{\frac{\pi}{4}} 
& \text{ if } m_1=0, m_2=1 
\\ 
-R_y(\theta_3+\pi)
R_x\left(\theta_2(m_1)\right) 
R_y(\theta_1) Z_{\frac{\pi}{4}} 
& \text{ if } m_1=1, m_2=0
\\
R_y(\theta_3)
\sigma_z
R_x\left(\theta_2(m_1)\right) 
R_y(\theta_1) Z_{\frac{\pi}{4}} 
& \text{ if } m_1=1, m_2=1
\end{cases} 
\\
&= 
\left(-1\right)^{m_3}
Z_{\frac{\pi}{4}} \sigma_z (\sigma_y)^{m_3 \oplus 1} \times 
\begin{cases}
- \sigma_z R_y(-\theta_3-\pi)
R_x\left(\theta_2(m_1)\right) 
R_y(\theta_1) Z_{\frac{\pi}{4}} 
& \text{ if } m_1=0, m_2=0 
\\ 
R_y(\theta_3)
R_x\left(\theta_2(m_1)\right) 
R_y(\theta_1) Z_{\frac{\pi}{4}} 
& \text{ if } m_1=0, m_2=1 
\\ 
- R_y(\theta_3+\pi)
R_x\left(\theta_2(m_1)\right) 
R_y(\theta_1) Z_{\frac{\pi}{4}} 
& \text{ if } m_1=1, m_2=0
\\
\sigma_z R_y(-\theta_3)
R_x\left(\theta_2(m_1)\right) 
R_y(\theta_1) Z_{\frac{\pi}{4}} 
& \text{ if } m_1=1, m_2=1.
\end{cases} 
\end{split}
\end{align}
As before, the modifications to $\theta_3$ can be performed with only knowledge of $m_1$ and $m_2$. Let $\theta_3\left(m_2,m_1\right)$ be defined as in the four cases of Eq. \ref{eq:cases2}, such that $\theta_3\left(m_2,m_1\right) = (-1)^{m_2 \oplus m_2 \oplus 1} \left(\theta_3 + \pi(1-m_2) \right)$. We perform the final measurement $m_3$ using this adaptive $\theta_3$. We obtain an equation of the desired form with a possible Pauli error term $\varepsilon(m_1, m_2, m_3)$ at the front:
\begin{align}
\begin{split}
U &= 
\left(-1\right)^{m_3}
Z_{\frac{\pi}{4}} \sigma_z (\sigma_y)^{m_3 \oplus 1} 
(-1)^{m_2} \sigma_z^{m_2 \oplus m_1 \oplus 1}
R_y\left(\theta_3(m_2, m_1)\right) 
R_x\left(\theta_2(m_1)\right) 
R_y(\theta_1) Z_{\frac{\pi}{4}} 
\\
&=
\left(-1\right)^{m_3 \oplus m_2 \oplus 1}
Z_{\frac{\pi}{4}} \sigma_z^{m_2 \oplus m_1} (\sigma_y)^{m_3 \oplus 1} 
R_y\left(\theta_3(m_2, m_1)\right) 
R_x\left(\theta_2(m_1)\right) 
R_y(\theta_1) Z_{\frac{\pi}{4}} 
\\
&=
\left(-1\right)^{m_3 \oplus m_2 \oplus 1}
\sigma_z^{m_2 \oplus m_1} (-\sigma_y)^{m_3 \oplus 1} 
Z_{\frac{\pi}{4}}
R_y\left(\theta_3(m_2, m_1)\right) 
R_x\left(\theta_2(m_1)\right) 
R_y(\theta_1) 
Z_{\frac{\pi}{4}} 
\\
& \equiv
\varepsilon(m_1, m_2, m_3)
Z_{\frac{\pi}{4}}
R_y\left(\theta_3(m_2, m_1)\right) 
R_x\left(\theta_2(m_1)\right) 
R_y(\theta_1) 
Z_{\frac{\pi}{4}},
\end{split}
\end{align}
where the error term $\varepsilon(m_1, m_2, m_3)$ is:
\begin{align}
\begin{split}
\varepsilon(0,0,0) &= -\sigma_y \\
\varepsilon(0,0,1) &= -\id \\
\varepsilon(0,1,0) &= -i \sigma_x \\
\varepsilon(0,1,1) &= \sigma_z \\
\varepsilon(1,0,0) &= -i \sigma_x \\
\varepsilon(1,0,1) &= \sigma_z \\
\varepsilon(1,1,0) &= -\sigma_y \\
\varepsilon(1,1,1) &= -\id.
\end{split}
\end{align}
We can remove any of these errors up to a global phase by using a sequence of non-interacting passes, where the atom is initialized to $\ket{g_0}$ or $\ket{g_1}$ rather than $\ket{+}$. To remove $-i \sigma_x$, we use a $\ket{g_0}$ initialization to apply $Z_\frac{\pi}{4} B B Z_\frac{\pi}{4} = i \sigma_x$. To remove $\sigma_z$, we use two $\ket{g_1}$-initialized scatterings to apply $Z_\frac{\pi}{4} B \sigma_z B Z_\frac{\pi}{4} Z_\frac{\pi}{4} B \sigma_z B Z_\frac{\pi}{4} = -i \sigma_z$. To remove $\sigma_y$, we apply two $\ket{g_1}$-initialized scatterings and one $\ket{g_0}$-initialized scatterings to apply  $Z_\frac{\pi}{4} B \sigma_z B Z_\frac{\pi}{4} Z_\frac{\pi}{4} B \sigma_z B Z_\frac{\pi}{4} Z_\frac{\pi}{4} B B Z_\frac{\pi}{4} = -i \sigma_y$. Thus, one can apply arbitrary single-qubit operations parameterized via $YXY$ Euler angles using this gate construction method.

\section{Photonic qubit readout}
\label{sec:swap_gate}

\begin{figure}[ht]
    \centering 
    \includegraphics[width=.6\textwidth]{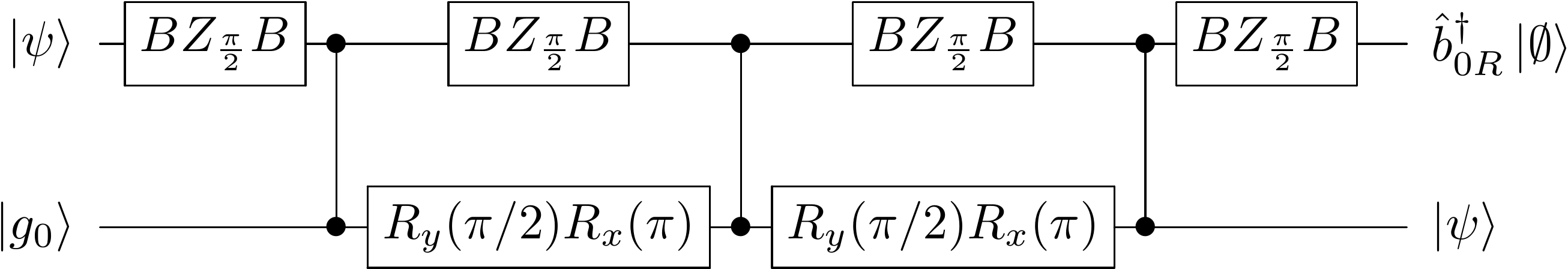}
    \caption{Construction of a SWAP gate from three scattering interactions. The top rail denotes the photonic qubit and the bottom rail denotes the atom. The $B Z_\frac{\pi}{2} B$ operations correspond to a return trip of the photon from the scattering site to the ring and back, passing through the beamsplitter and phase shifter twice.}
    \label{fig:swap_circuit}
\end{figure}

To measure the state of a photonic qubit, we construct a SWAP gate from a sequence of three scattering operations. We may initialize the atom to any state, then perform the sequence of scattering interactions shown in Figure \ref{fig:swap_circuit}. 

Let $\ket{\psi} = \left(\alpha \bdag_{0L} + \beta \bdag_{1L} \right)\vac$ be the state of the photon at points $P_3, P_4$ in the device. By scattering the photon against the atom three times and applying the rotation $R_y (\pi/2) R_x (\pi)$ to the atomic states in between scattering, one can swap the states of the photon and atom, such that the final atomic state is $\alpha \ket{g_0} + \beta \ket{g_1}$. It is straightforward to verify that this sequence of operations implements the SWAP gate up to a phase of -1:
\begin{equation}
\left(B Z_{\frac{\pi}{2}} B \otimes \id \right) \cz
\left(B Z_{\frac{\pi}{2}} B \otimes Y_{\frac{\pi}{2}} X_\pi \right) \cz
\left(B Z_{\frac{\pi}{2}} B \otimes Y_{\frac{\pi}{2}} X_\pi \right) \cz
\left(B Z_{\frac{\pi}{2}} B \otimes \id \right) = 
-1 \begin{pmatrix}
1 & 0 & 0 & 0 \\
0 & 0 & 1 & 0 \\
0 & 1 & 0 & 0 \\
0 & 0 & 0 & 1
\end{pmatrix}.
\end{equation}

Once the states of the photonic and atomic qubits are swapped, the atomic state can be measured with near 100\% efficiency using the quantum jump technique \cite{Monroe2002QuantumCavities, Duan2004ScalableInteractions} while the photonic qubit is discarded by allowing it to gradually dissipate through leakage to the environment. This SWAP-and-measure protocol can be repeated for the rest of the photonic qubits to read out the entire photonic quantum state.

\section{Implementing a two-photon $\cz$ gate}
\label{sec:control_z}

In addition to implementing single-qubit gates, constructing a two-photon entangling gate is necessary for universal computation. A controlled phase-flip gate $\cz$ between two photonic qubits can be constructed through a sequence of three scattering interactions in a somewhat similar manner as in Ref. \cite{Duan2004ScalableInteractions}. However, the fixed beamsplitter and phase shifter, which are required for implementation of single-qubit gates in our scheme, only allow us to apply operations of the form $\left( (Z_\frac{\pi}{4} B) \otimes \id \right) \cz \left( (B Z_\frac{\pi}{4}) \otimes \id \right)$ to the $\ket{\text{photon}} \otimes \ket{\text{atom}}$ system with each scattering interaction. This prevents us from performing the exact protocol described in Ref. \cite{Duan2004ScalableInteractions}, which requires photons to undergo three successive $\cz$ operations without any gates between them.

Here we describe two possible implementations of this $\cz$ gate between two photons $A$ and $B$ in states $\ket{\psi_A}$ and $\ket{\psi_B}$ which work with the design of our proposed device. The first solution is to use a SWAP gate as described in Section \ref{sec:swap_gate} to swap the states of photon $A$ and the atom, then perform a scattering of photon $B$ against the atom, then to swap the atomic state back to photon $A$. 

Although the construction of $\cz$ through SWAP gates allows for direct interaction of $\ket{\psi_A}$ with $\ket{\psi_B}$, it involves a total of $3+1+3=7$ scattering interactions, which is significantly less compact than the three scatterings used in the construction from Ref. \cite{Duan2004ScalableInteractions}. 

\begin{figure}[ht]
\centering
\includegraphics[width=\textwidth]{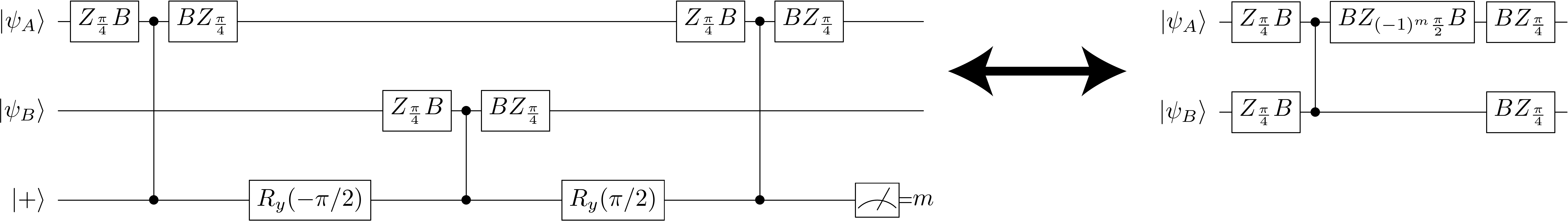}
\caption{Construction of a $\cz$ gate with three scattering interactions using a measurement-based approach. After measurement, the left and right circuits are equivalent. The single-qubit gates on either side of $\cz$ can be removed by absorbing them into the preceding/subsequent single-qubit gates as described above.}
\label{fig:cz2}
\end{figure}

We can implement a more compact construction of $\cz$ which also only requires three scatterings by using a measurement based scheme shown in Figure \ref{fig:cz2}. This second possible construction implements a $\cz$ gate between photons $A$ and $B$ which is sandwiched between single-qubit gates. These extra gates can be implicitly removed by programming the single-qubit gate $U_\text{before}$ which immediately precedes this operation to instead implement $\left(Z_\frac{\pi}{4} B\right)^{-1} U_\text{before}$ and the gate $U_\text{after}$ following $\cz$ to implement $U_\text{after} \left(B Z_\frac{\pi}{4} \right)^{-1}$.

\section{Circuit compilation}
\label{sec:circuit_compilation}

\begin{figure}[ht]
\centering
\includegraphics[width=\textwidth]{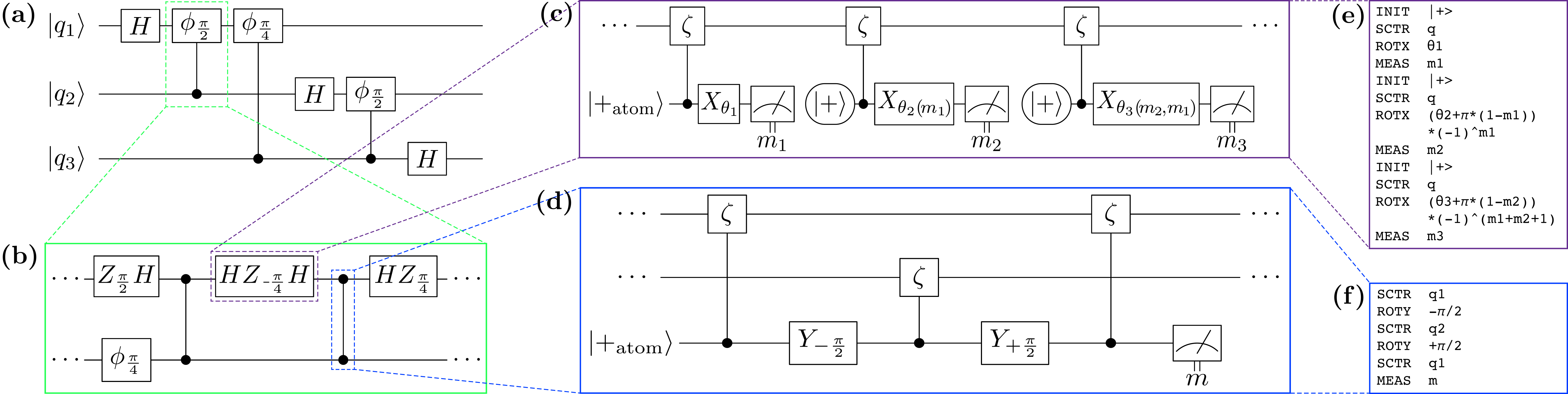}
\caption{
Graphical depiction of the circuit compilation process. 
\textbf{(a)} The target quantum circuit we wish to implement in the device, in this case a three-qubit quantum Fourier transform. 
\textbf{(b)} The first step of the compilation process is to decompose complex circuit elements into single-qubit and $\cz$ gates. The subcircuit depicted here implements the first controlled-$\phi_\frac{\pi}{2}$ gate between photonic qubits $q_1$ and $q_2$. 
\textbf{(c, d)} The second step is to decompose each single-qubit gate (c) via Euler angles into a sequence of rotations which can be teleported from the atom to the photonic qubits, and to decompose each $\cz$ gate (d) using the scattering sequence shown in Figure \ref{fig:cz2}.
\textbf{(e, f)} Programmatic representation of the instructions sent to the device to implement subroutines (c,d), respectively. The full code for implementing the target quantum circuit depicted in (a) is shown in Program \ref{prg:qft_code}.
}
\label{fig:compilation}
\end{figure}

An arbitrary $n$-qubit quantum operator $U \in \mathrm{U}(2^n)$, can be compiled into a sequence of physical instructions on the proposed device using a three-step process shown in Figure 4 of the main text, and shown in greater detail in Figure  \ref{fig:compilation} of this document. The first step is to decompose $U$ into a sequence of single-qubit gates and $\cz$ operations, a process described in our previous work \cite{Bartlett2020UniversalProcessing}. The second step is to decompose each single-qubit gate via Euler angles as three $R_y$ rotations which may be teleported onto the photonic qubits by a sequence of scatter-rotate-measure operations. The third step is to use a high-speed classical control system to modify the adaptive rotations which are applied to the atomic qubit based on the measurement outcomes during operation. Pauli errors which are accumulated during the course of the circuit operation can either be removed explicitly by scattering against $\ket{g_0}$ or $\ket{g_1}$, as described at the end of Section \ref{sec:single_qubit_gates}, or can be removed implicitly (resulting in a more compact circuit) by programming the inverse of the error term into subsequent single-qubit operators. An example program for implementing a three-qubit quantum Fourier transform is shown in Program \ref{prg:qft_code} at the end of this Supplementary Information document.

\section{Imperfection analysis}
\label{sec:imperfection_analysis}

Here we describe the details of the imperfection analysis that we used for estimating the achievable circuit depth, shown in Figure 4 of the main text. The main sources of error for our protocol are the same as for the Duan-Kimble protocol~\cite{Duan2004ScalableInteractions}, but with the added loss from the switches and propagation loss through the storage ring. We group these errors into three main classes:
\begin{itemize}
    \item Pulse shape infidelity: mismatch between the cavity output pulses for the atom being in the $\ket{g_0}$ and $\ket{g_1}$ states. This loss can be minimized by choosing the photon's temporal width ($\tau$)  to be much larger than the cavity photon lifetime $1/\kappa$: $\kappa \tau \gg 1$.
    \item Spontaneous emission loss of the excited state of the atom, where the atom in the $\ket{e}$ state emits not into the desired cavity mode but into other modes or into free space. In our scheme, this causes photon leakage error when the atom is in the $\ket{g_1}$ state, since the photon causes the $\ket{g_1}$ state to temporarily transition to $\ket{e}$.
    \item Photon loss due to optical elements. This includes optical attenuation while propagating through the storage ring, insertion loss of the optical switches, and spurious loss from the cavity mirrors or the cavity medium.
\end{itemize}

We assume that the cavity mode at $\omega_c$ is resonant with the atom $\ket{g_1} \leftrightarrow \ket{e}$ transition frequency $\Omega_1$, since the detuning can be actively tuned to be zero, both in free-space by tuning the cavity length, as well as in solid-state nanophotonic systems through temperature or strain. We also assume that rotations of the atomic state by the cavity laser and measurement of the state via the quantum jump technique can be done with fidelity $\mathcal{F} \approx 1$, since both processes have been demonstrated experimentally with very high fidelities~\cite{gehr_cavity-based_2010} greatly exceeding that of the effects listed above.

To quantify the effects of these sources of error, we assume the input waveguide contains a single photon Fock state of the form $\int dt\, \phi_{\rm in}(t)\, \hat a_{\rm in}^\dagger(t) \vac$, where $\phi_{\rm in}(t)$ is the pulse shape, $\vac$ represents the vacuum state of the waveguide modes, and $\hat a_{\rm in}^\dagger(t)$ is a bosonic operator obeying the standard commutation relation $[\hat a_{\rm in} (t), \hat a_{\rm in}^\dagger (t')] = \delta(t-t')$ which creates a photon propagating toward the cavity in the waveguide at time $t$. For the cavity output, we assume a similar form, $\int dt\, \phi_{\rm out}(t)\, \hat a_{\rm out}^\dagger(t) \ket{\varnothing}$~\cite{Duan2004ScalableInteractions, Shen2009TheoryAtomb}, where $\hat a_{\rm out}^\dagger(t)$ is similarly defined and creates a photon propagating away from the cavity at time $t$. For our analysis, we choose a Gaussian pulse envelope centered at $t_0=\Delta t/2$ for the input: $\phi_{\rm in}(t) \propto \exp\left[-( t-t_0)^2/\tau^2\right]$, as studied in Ref.~\onlinecite{Duan2004ScalableInteractions}. 

To solve for the output single-photon pulse, we use the analytical technique described by Shen and Fan~\cite{Shen2009TheoryAtom, Shen2009TheoryAtomb}, which exactly solves the single-photon transport problem of a coupled atom-cavity-waveguide system, taking into account all relevant energy scales. 
The effective Hamiltonian of the overall system is given by~\cite{Shen2009TheoryAtomb}:

\begin{align}
\begin{split}
\label{eq:hamiltonian}
\mathcal{H_{\rm eff}}/\hbar = &  
\left(\omega_c - i\kappa_{\rm i}/2\right) \, \hat a^\dagger \hat a + (\Omega_e-i\gamma_s/2) \ketbra{e}{e} + \Omega_1 \ketbra{g_1}{g_1} + \Omega_0 \ketbra{g_0}{g0} + \left( g\, \hat a^\dagger \ketbra{g_1}{e} + {\rm H.c.}\right )\\
&+ \int dx\, \delta(x) \left[ \sqrt{\kappa v_{\rm g}/2}\ \hat a^\dagger \hat a_{\rm in}(x)  + \sqrt{\kappa v_{\rm g}/2}\ \hat a^\dagger \hat a_{\rm out}(x) + {\rm H.c.}\right] \\
&+ \int dx\, \hat a_{\rm in}^\dagger(x) (\omega_c - iv_{\rm g} \partial_x) \,\hat a_{\rm in}(x) + \int dx\, \hat a_{\rm out}^\dagger(x) (\omega_c + iv_{\rm g} \partial_x) \,\hat a_{\rm out}(x),
\end{split}
\end{align}
where $\hat a^\dagger$ is a bosonic operator that creates a photon in the cavity mode at $\omega_c$ obeying $[\hat a, \hat a^\dagger] = 1$, $\kappa_{\rm i}$ is the intrinsic dissipation rate of the cavity mode, $\Omega_{\rm 0, 1, e}$ are the energies of the respective atomic states, $g$ is the single-photon atom-cavity coupling rate (equal to half the vacuum Rabi splitting), $v_{\rm g}$ is the group velocity of the waveguide in the vicinity of the cavity resonant frequency $\omega_c$, and $\gamma_s$ is the spontaneous emission rate of the atomic $\ket{e}$ state\footnote{
One should note that, while the use of the non-Hermitian $-i \gamma_s / 2 \ketbra{e}{e}$ term is known to produce correct scattering matrices for single-photon interactions, the direct substitution of $\Omega_e \rightarrow \Omega_e - i \gamma_s / 2$ to describe spontaneous emission loss will yield incorrect results for temporally-overlapping multi-photon scattering interactions.~\cite{Rephaeli2013DissipationInvited} The more correct treatment here is to add additional couplings between the system Hamiltonian and a bath of modes describing the environment, but this is not necessary for our analysis, which is limited to single-photon interactions.
}. In the following analysis, we set $\kappa_{\rm i} = 0$.

The spectrum of the output pulse, $\tilde \phi_{\rm out}(\omega) =\mathcal{F}\{\phi_{\rm out} (t)\}$ is related to the input pulse spectrum $\tilde \phi_{\rm in}(\omega) = \mathcal{F}\{\phi_{\rm in} (t)\}$ by the spectral response of the cavity-atom system $R(\omega, g, \kappa, \gamma_s, \ket{A})$. Here, $\mathcal{F\{\cdot\}}$ denotes the Fourier transform, and $\omega$ denotes the input photon detuning from the cavity/atom resonance, $\omega = \omega_{\rm in} - (\Omega_e-\Omega_1) = \omega_{\rm in} - \omega_c$. 
The spectral response depends on the initial state of the atom $\ket{A} \in \{\ket{g_0}, \ket{g_1}\}$. This treatment captures the full quantum mechanical response of the system to a single-photon Fock state input for an arbitrary initialization of the atom, without making the semiclassical assumption of a weak coherent state for the input.

\emph{Pulse shape infidelity and delay correction ---}
For an atom initialized as $\ket{A} = \ket{g_0}$, the response is identical to an empty cavity since the $\ket{g_0} \leftrightarrow \ket{e}$ transition frequency is far-detuned from the cavity mode frequency $\Omega_c$~\cite{Duan2003CavityAtoms}. In this case, the output pulse is slightly delayed from the input pulse by a time $\delta t_0$, as it couples into the empty cavity mode before coupling out, leading to a fidelity below unity, as shown in Figure 4 of the main text. For an initialization $\ket{A} = \ket{g_1}$, the photon is directly reflected from the front mirror of the cavity, since the dressed cavity modes are well-separated from the input photon frequency by the vacuum Rabi splitting for strong coupling $g \gg \kappa, \gamma_s$, and the delay $\delta t_1 \approx 0$ is minimal.  Here the pulse shape fidelity is defined as~\cite{Duan2003CavityAtoms, duan_robust_2005}:
\begin{equation}
    \mathcal{F}_\text{shape} \equiv \left| \int dt\, \bar\phi^*_{\rm in} (t) \, \bar\phi_{\rm out}(t) \right|,
\end{equation}
where $\bar\phi_{\rm in}$ and $\bar\phi_{\rm out}$ are the renormalized input and output pulses. The pulse shape infidelity is defined as $1-\mathcal{F}_\text{shape}$. Importantly, this quantity only describes the infidelity due to shape mismatch of the input and output pulses, not amplitude mismatch; the infidelity due to spontaneous emission loss is computed separately. The average infidelity for an initialization in the $\ket{+} = ({\ket{g_0} + \ket{g_1}})/\sqrt{2}$ state is calculated as the mean of the infidelities for the $\ket{g_0}$ and $\ket{g_1}$ states~\cite{Duan2004ScalableInteractions}. In our calculations, using a long pulse width $\tau = 100/\kappa$ and total interaction timescale $T=500/\kappa$ and assuming no intrinsic losses in the cavity ($\kappa_{\rm i} = 0$) aside from spontaneous emission results in a low infidelity below $10^{-3}$ per photon-cavity scattering event. 

In Figure 4(b) of the main text, we plot the shape infidelity of various states as a function of the single-atom cavity cooperativity $C \equiv 4g^2 / \kappa \gamma_s$, where $\gamma_s$ measures the spontaneous emission rate and is fixed at $\gamma_s=\kappa/5$. The pulse shape infidelity of an interaction with the $\ket{g_1}$ state decreases to negligible values as $C$ increases, while the infidelity of $\ket{g_0}$ reaches an asymptote at $8\times 10^{-4}$ due to the delay of the output pulse by a time $\delta t_0$ which is independent of $C$; the infidelity of the $\ket{+}$ interaction asymptotes at $4 \times 10^{-4}$. Since the atom will usually be initialized to the $\ket{+}$ state during operation of the device, it is useful to minimize the infidelity of interacting with this state. This can be done by delaying the reference pulse by a time difference $t_\text{delay} = (\delta t_0 + \delta t_1) / 2 \approx \delta t_0 / 2$ by adding an additional path length $c\, t_\text{delay} / 2$ to the top waveguide in Figure 1 of the main text. This distributes the infidelity due to the output pulse delay equally between the $\ket{g_0}$ and $\ket{g_1}$ states, such that the output pulse of a $\ket{g_1}$ interaction is shifted forward by $\delta t_0/2$ and the output of a $\ket{g_0}$ interaction is delayed by $\delta t_0/2$. This results in an infidelity of approximately $2 \times 10^{-4}$ which is independent of both cavity cooperativity (at $C \gg 1$) and atomic state initialization.

\emph{Spontaneous emission loss ---}
Atomic spontaneous emission noise from the excited $\ket{e}$ state at a rate $\gamma_s$ results in a partial loss of the photon, resulting in an output pulse with total photon number $\int dt\, |\phi_{\rm out}(t)|^2 < 1$. We calculate the probability $P_s$ of spontaneous emission loss as:
\begin{equation}
    P_s = 1 - \frac{\int dt\, |\phi_{\rm out}(t) |^2 }{\int dt\, |\phi_{\rm in}(t)|^2}.
\end{equation}
Spontaneous emission noise only applies to the $\ket{1} \otimes \ket{g_1}$ component of the $\text{photon} \otimes \text{atom}$ state. The atom will usually be initialized to the $\ket{+}$ state, and averaging over possible input photon states, we obtain an average leakage probability of $\bar P_s = P_s/4$, as shown in Figure 4(b), which is well-approximated by $\bar P_s = [4(1+2C)]^{-1}$.

\emph{Spurious photon loss and maximum circuit depth ---}
Finally, we account for loss due to propagation through the optical paths and switches as an average loss per cycle $L$. To estimate the maximum circuit depth $D$ attainable with an overall fidelity $\mathcal{F} > \mathcal{F}_\text{target}$, we compute a ``bulk fidelity'' accounting for shape mismatch and loss due to average spontaneous emission and propagation through the storage ring. For simplicity, we assume the circuit operates on only a single photonic qubit and that the photon is scattered off the atom with every pass through the storage ring. The achievable circuit depth operating with success probability $P_\text{success} = \mathcal{F}_\text{target}$ is thus the maximum $D$ satisfying:
\begin{equation}
    \left [\mathcal{F}_\text{shape} \times (1 - \bar P_s) \times (1 - L) \right]^{D} \ge \mathcal{F}_\text{target},
\end{equation}
which is plotted as a function of cavity cooperativity and propagation loss in Figure 4(c) in the main text.

\newpage

\begin{code}
\twocolumngrid
\begin{minted}[
	bgcolor=gray!6,
	fontfamily=tt,
	gobble=0,
	fontsize=\small,
	breaklines=true,
	frame=lines,
	framesep=1mm,
	funcnamehighlighting=true,
	tabsize=4,
	obeytabs=true,
	mathescape=false
	samepage=true,
	showspaces=false,
	showtabs =false,
	texcl=false,
	linenos=true,
	numberblanklines=true,
	xleftmargin=1pt,
	numbersep=1pt
]{text}
# Instruction set
# ---------------
# OPEN t ... open the switches at time t
# CLOS t ... close the switches at time t
# ROTX θ ... laser pulse rotates atom state, Rx(θ)
# ROTY θ ... laser pulse rotates atom state, Ry(θ)
# MEAS m ... measure atom state and store bit as m
# INIT Ψ ... initialize atom to |Ψ>=|g0>,|g1>,|+>


# Scatter photon q and return it to ring
define SCTR q:
	OPEN  t_q-Δt/2        # t_q: time bin for |q>
	CLOS  t_q+Δt/2        # Δt: temporal bin size
	OPEN  N*Δt+t_q-Δt/2   # N: number of time bins
	CLOS  N*Δt+t_q+Δt/2   # N*Δt: time around ring


# Explicitly correct Pauli errors after a gate
define CORR q m1 m2 m3:
	if m3 == 0:
		INIT  |g1>
		SCTR  q
		SCTR  q
		INIT  |g0>
		SCTR  q
	if m1 != m2:
		INIT  |g1>
		SCTR  q
		SCTR  q


# Single-qubit gate via Euler angles
define GATE q θ1 θ2 θ3:
	INIT  |+>
	SCTR  q
	ROTX  θ1
	MEAS  m1
	INIT  |+>
	SCTR  q
	ROTX  (θ2+π*(1-m1))*(-1)^m1   # adaptive θ2
	MEAS  m2 
	INIT  |+>
	SCTR  q
	ROTX  (θ3+π*(1-m2))*(-1)^(m1+m2+1)
	MEAS  m3
	CORR  q m1 m2 m3   # remove Pauli ε(m1,m2,m3)


# Swap photon q with atom state
define LOAD q:
	SCTR  q
	ROTX  π
	ROTY  π/2
	SCTR  q
	ROTX  π
	ROTY  π/2
	SCTR  q
	ROTX  π/2
	ROTY  π/4


# Controlled-σz between photons q1, q2
define CTRZ q1 q2:
	GATE  q1  0  3π/4  -π/2
	GATE  q2  0  3π/4  -π/2
	SCTR  q1
	ROTY  -π/2
	SCTR  q2
	ROTY  +π/2
	SCTR  q1
	MEAS  m
	GATE  q1  m*π  π/2   (-1)^m*3π/2
	GATE  q2  π/2  3π/4  0


# Run a 3-qubit QFT and measure the qubits
GATE  q1  5.668  2.094  0.615   # H
GATE  q1  3.757  2.094  5.668   # cφ(π/2)
CTRZ  q2  q1  
GATE  q1  2.101  1.718  4.182
CTRZ  q2  q1  
GATE  q1  0.000  2.356  1.571
GATE  q3  1.571  0.785  4.712   # cφ(π/4)
GATE  q1  4.712  2.356  0.000
CTRZ  q3  q1  
GATE  q1  1.845  1.609  4.438
CTRZ  q3  q1  
GATE  q1  5.918  2.283  1.041
GATE  q2  5.668  2.094  0.615   # H
GATE  q2  3.757  2.094  5.668   # cφ(π/2)
CTRZ  q3  q2  
GATE  q2  2.101  1.718  4.182
CTRZ  q3  q2  
GATE  q2  0.000  2.356  1.571
GATE  q2  5.668  2.094  0.615   # H

# State readout
LOAD  q1
MEAS  b1
LOAD  q2
MEAS  b2
LOAD  q3
MEAS  b3
\end{minted}
\caption{Assembly-like pseudocode for implementing the three-qubit quantum Fourier transform shown in Figure \ref{fig:compilation}a. For simplicity and readability, this code explicitly corrects for Pauli errors using the \texttt{CORR} subroutine and removes extraneous $B Z$ terms in the $\cz$ gate construction using four additional \texttt{GATE} calls within \texttt{CTRZ}. The numerical values for the \texttt{GATE} angles in lines 78-96 were computed using a modified version of \texttt{OneQubitEulerDecomposer} in \texttt{Qiskit} \cite{Qiskit}.}
\label{prg:qft_code}
\end{code}

\onecolumngrid

\newpage


%